\documentclass{aa}

\usepackage[utf8]{inputenc}
\usepackage[T1]{fontenc}

\usepackage[english]{babel}
\usepackage{txfonts}
\usepackage{charter}
\usepackage{balance}
\usepackage{hyperref}
    \hypersetup{colorlinks=true,linkcolor=blue,citecolor=blue,urlcolor=blue}
    \makeatletter
    \renewcommand*\aa@pageof{, page \thepage{} of \pageref*{LastPage}}
    \makeatother

\usepackage{siunitx}

\usepackage{booktabs}
\usepackage[flushleft,para]{threeparttable}
\usepackage{multirow}

\usepackage[nowarn]{glossaries}
    \makeglossaries

\newacronym{1D}{1D}{1-dimensional}

\newacronym{BJD}{BJD}{Barycentric Julian Date}

\newacronym{CB}{CB}{convective blueshift}

\newacronym{FAP}{FAP}{false alarm probability}

\newacronym{GLS}{GLS}{generalized Lomb-Scargle}

\newacronym{HARPS}{HARPS}{High Accuracy Radial velocity Planet Searcher}
\newacronym{HARPS-N}{HARPS-N}{HARPS-North}
\newacronym{HMI}{HMI}{Helioseismic and Magnetic Imager}

\newacronym{KS}{KS}{Kolmogorov-Smirnov}

\newacronym{LBL}{LBL}{line-by-line}
\newacronym{LTE}{LTE}{local thermodynamic equilibrium}

\newacronym{RMS}{RMS}{root mean square}
\newacronym{RV}{RV}{radial velocity}

\newacronym{SDO}{SDO}{Solar Dynamics Observatory}
\newacronym{S/N}{S/N}{signal-to-noise ratio}
\newacronym{SME}{SME}{Spectroscopy Made Easy}
\newacronym{SOAP}{SOAP}{Spot Oscillation And Planet}

\newacronym{VALD}{VALD}{Vienna Atomic Line Database}

\def\aCenB{$\alpha\,\mbox{Cen}\,\mbox{B}$}

\begin{document}

   \title{Measuring precise radial velocities on individual spectral lines}
   \subtitle{III. Dependence of stellar activity signal on line formation temperature}

   \author{K. Al Moulla\inst{1}
          \and X. Dumusque\inst{1}
          \and M. Cretignier\inst{1}
          \and Y. Zhao\inst{1}
          \and J. A. Valenti\inst{2}
          }

   \institute{Observatoire Astronomique de l'Université de Genève, Chemin Pegasi 51, 1290 Versoix, Switzerland\\
              \email{khaled.almoulla@unige.ch}
              \and Space Telescope Science Institute, 3700 San Martin Drive, Baltimore, MD 21218, USA
             }

   \date{Received 6 February 2022 / Accepted 9 May 2022}

\abstract
{To enable \acrfull*{RV} precision on the order of ${\sim}\SI{0.1}{\meter/\second}$ required for the detection of Earth-like exoplanets orbiting solar-type stars, the main obstacle lies in mitigating the impact of stellar activity.}
{This study investigates the dependence of derived RVs with respect to the formation temperature of spectral line segments.}
{Using spectral synthesis, we compute for each observed wavelength point of unblended spectral lines the stellar temperature below which $\SI{50}{\percent}$ of the emergent flux originates. We can then construct RV time series for different temperature ranges, using template matching.}
{With HARPS-N solar data and HARPS \aCenB{} measurements, we demonstrate on time intervals of prominent stellar activity that the activity-induced RV signal has different amplitude and periodicity depending on the temperature range considered. We compare the solar measurements with simulated contributions from active surface regions seen in simultaneous images, and find that the suppression of convective motion is the dominant effect.}
{From a carefully selected set of spectral lines, we are able to measure the RV impact of stellar activity at various stellar temperatures ranges. We are able to strongly correlate the effect of convective suppression with spectral line segments formed in hotter temperature ranges. At cooler temperatures, the derived RVs exhibit oppositely directed variations compared to the average RV time series and stronger anti-correlations with chromospheric emission.}

\keywords{
stars: activity -- stars: individual: \object{Sun} -- stars: individual: \object{HD128621} -- techniques: radial velocities -- techniques: spectroscopic}

\authorrunning{K. Al Moulla et al.}
\titlerunning{Measuring precise radial velocities on individual spectral lines. III.}

\maketitle


\section{Introduction}\label{Sect:1}

In the last $25$ years, the \acrfull*{RV} technique has been highly successful in the search of planets orbiting other stars. Technical advancements have enabled the discovery of a wide range of exoplanetary types, in stellar systems with architectures far different from our own Solar System. However, the strive to find smaller, Earth-like planets is currently impeded by stellar variability drowning out potential planetary RV signals. The precision required to detect an Earth twin around a solar-like star is of the order of ${\sim}\SI{0.1}{\meter/\second}$, about one order of magnitude smaller than present-day attainability using the best activity mitigation techniques \citep[e.g.][]{Zhao+22}.

By cross-correlating an observed stellar spectrum with a template of the laboratory wavelengths and expected strengths of absorption lines (depending on the spectral type of the star), RVs have traditionally been derived using the entire spectrum at once. Previous studies \citep{Dumusque18,Cretignier+20a} show that a careful selection of \acrfull*{LBL} \acrshortpl*{RV} can help mitigate the effects of stellar activity. \cite{Cretignier+20a} showed that the RV of some lines are uncorrelated with respect to the mean RV of all available lines, and thus seem to be less affected by stellar activity.

The effects of stellar variability on RVs manifest themselves in many different ways. In the structure of low-mass stars, convection in the envelope leads to stellar oscillations and granulation patterns on the stellar surface. Stellar oscillations are due to the propagation of acoustic waves in the convective envelope of late-type stars, which locally modify the stellar surface and can be measured using RVs \citep[e.g.][]{Kjeldsen&Bedding95,Arentoft+08}. The granulation patterns, consisting of ascending hot gas cells and descending cool ones, form blueshifted granules and redshifted intergranulation lanes. Considering the spectral line contributions across a stellar disk, a component formed at the location of a granule would be slightly blueshifted and deeper (due to the higher intensity of granules) than if observed elsewhere \citep{Nordlund+09}. Due to a larger coverage in area and a higher brightness, the granules lead to an observed net \acrfull*{CB} of disk-integrated lines, as well as asymmetries induced by the weaker redshifted contributions. These shifts and asymmetries, known as the third signature of granulation \citep{Gray09}, differ for different spectral lines, with weak lines being affected the most \citep{Reiners+16,Meunier+17a}.

However, stellar oscillations and the variation of the granulation pattern induce RV perturbations on timescales shorter than a few days. These phenomena are therefore not responsible for the dominant RV perturbations which occur on timescales of several days and longer. In fact, both surface oscillations and granulation have been shown to be relatively easy to mitigate with the choice of an appropriate observation strategy \citep{Dumusque+11,Meunier+17b}. This is done by choosing a longer exposure time and observing the target star multiple times per night with sufficient temporal gaps. On timescales longer than a few days, the dominant RV signal is induced by the impact of active regions, which are stellar surface features caused by concentrated magnetic fields \citep{Schrijver&Zwaan:2000}. These active regions can either be spots, which have a lower contrast compared to the inactive stellar surface, or faculae, which have a slightly higher contrast.

Active regions influence the RV precision in mainly two ways \citep{Bauer+18}. The first effect arises because spots and faculae break the flux balance between the two hemispheres which have Doppler velocities of opposite signs due to rotation. This introduces quasi-sinusoidal variations with a periodicity of half the stellar rotation, imitating the Rossiter-McLaughlin effect caused by a transiting planet, but somewhat dissimilar because of spherical projection and varying filling factor at any given time as spots and faculae grow and decay while moving in and out of sight. The second effect arises when the same magnetically active regions inhibit underlying convective motion, which introduces a modulation with a period of the stellar rotation. \cite{Meunier+10} showed that the suppression of convective blueshift, which unlike the flux effect cannot balance itself out when the approaching and receding hemispheres are equally covered, is the dominating contribution in solar simulations, reaching peak-to-peak variations of ${\sim}\SI{10}{\meter/\second}$.

This paper further investigates the selection of spectral lines used to derive RVs, applying a diagnostic tool obtained through spectral synthesis. We implement a novel approach in which line segments formed at different temperature ranges are believed to be more or less susceptible to stellar activity compared to their LBL averages. To test this, we explore whether RV perturbations are stronger in spectral intervals formed deeper in the photosphere, where radial flow velocities and temperature contrasts are larger.

In Sect.~\ref{Sect:2}, we describe the data that we use in the following analysis. In Sect.~\ref{Sect:3}, we describe our method of obtaining the average formation temperature of each sampled point of a stellar spectrum, and then explain how to measure the RV for different temperature ranges. We apply our method to the Sun and \aCenB{} in Sect.~\ref{Sect:4} and demonstrate that the stellar activity signal changes as a function of temperature in the stellar atmosphere. We finally discuss the obtained results in Sect.~\ref{Sect:5}.

\section{Data}\label{Sect:2}

\subsection{Observed spectra}\label{Sect:2.1}

In this paper, we analyse spectra of the Sun and \aCenB{} which have been obtained with HARPS-N and HARPS, respectively. Both instruments have very similar resolution, $R\,{=}\,115,000$. More precisely, we analyse the echelle-order merged spectra produced by the HARPS and HARPS-N data reduction software. Those spectra are interpolated on a common wavelength grid to facilitate the comparison of all spectra.

Since we are interested in the stellar activity signal that induces RV variations on the rotational period timescale, i.e. dozens of days for Sun-like stars, we daily bin the available spectra to increase the \acrfull*{S/N}, which will be beneficial to derive more precise information at the spectral line level. The median S/N values for the Sun and \aCenB{} daily-binned spectra are $890$ and $590$, respectively. We then correct for the color of those daily-binned spectra using \texttt{RASSINE} \citep{Cretignier+20b}, to get the same continuum level for all the spectra, and post-process them using \texttt{YARARA} \citep{Cretignier+21} to remove telluric contamination and known instrumental systematics. To reach the highest S/N possible, a master spectrum is built by combining all available observations for each target. We note that the master spectrum is shifted to the stellar rest frames, to be able to easily compare the spectral lines with line lists from the VALD3\footnote{VALD stands for \acrlong*{VALD}.} database \citep{Piskunov+95,Kupka+00,Ryabchikova+15}.

\subsection{Synthesised spectra}\label{Sect:2.2}

To obtain the average formation height in the stellar atmosphere for each sampled point of a spectrum, we take advantage of spectral synthesis to evaluate the formation contribution function. We perform \acrfull*{1D} \acrfull*{LTE} spectral synthesis using \texttt{PySME}\footnote{Available at \url{https://github.com/AWehrhahn/SME}.}, a Python implementation of \acrlong*{SME} \citep[SME;][]{Valenti&Piskunov96,Piskunov&Valenti17}. For modeling the spectra of the Sun and \aCenB{}, we use \texttt{MARCS} model atmospheres \citep{Gustafsson+08} with different stellar parameters and line lists from VALD3 where the line detection threshold is set to $\SI{1}{\percent}$ in accordance with the chosen parameters.

The synthesized spectra are computed on the same wavelength grid as the master spectra of the targets (see Sect.~\ref{Sect:2.1}). For each target, two synthesized spectra will be produced. The first will be generated at the instrumental resolution, and will include micro- and macroturbulences, and rotational broadening using the projected rotational velocity. The goal of this synthesized spectrum is to model as closely as possible the observed spectra. Another synthesis will be generated without turbulence and rotational broadening at the highest resolution possible for the observed wavelength grid. This synthesis will be used to detect line blends which are unresolved at the HARPS and HARPS-N resolution.

\begin{figure*}[t!]
	\includegraphics[width=\textwidth]{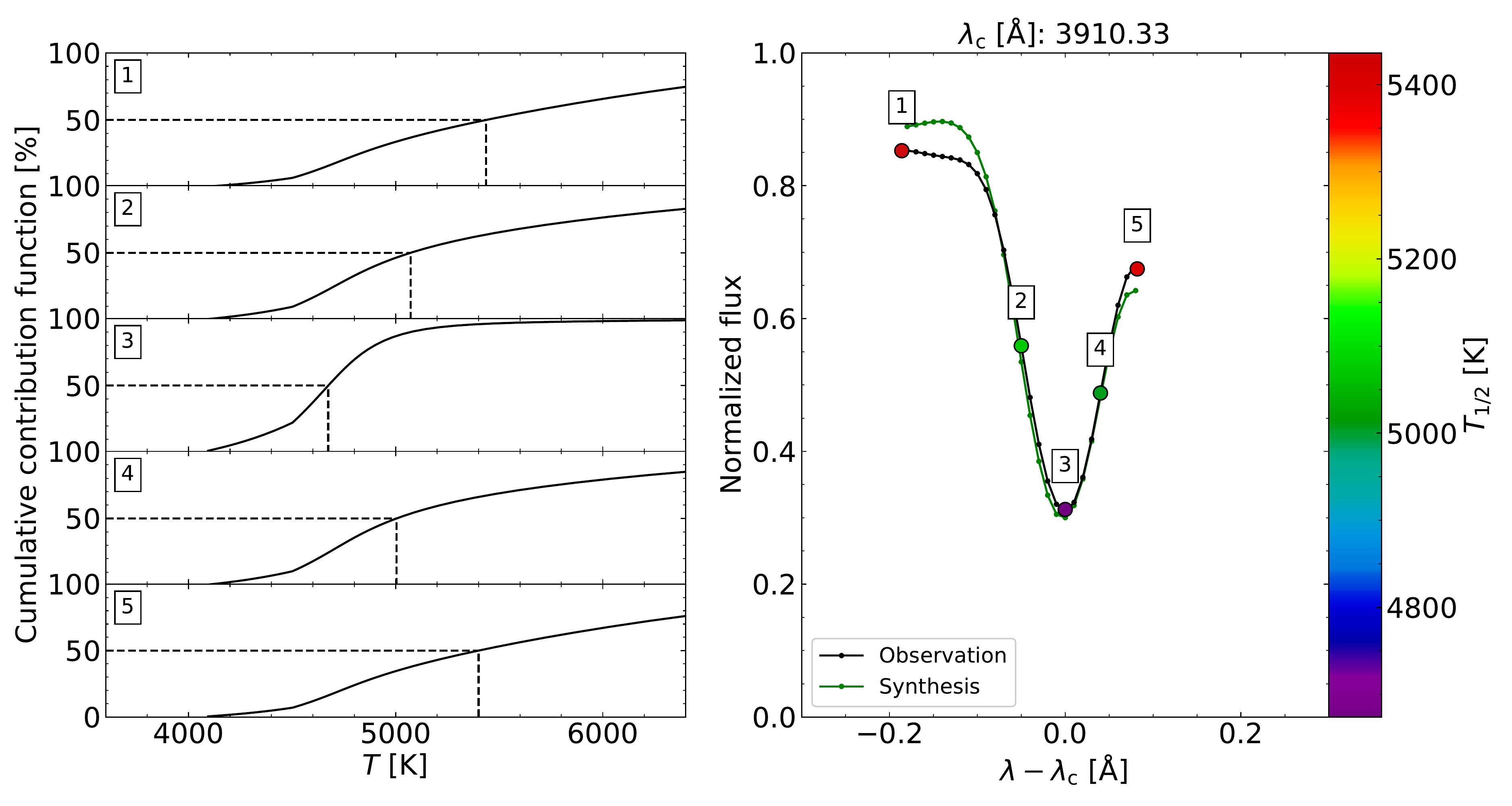}
	\caption{\textit{Left}: Normalized cumulative contribution function as a function of temperature at $5$ points across the spectral line shown in the right panel. The dashed line indicates the location of $T_{1/2}$, i.e. where the contribution function is $\SI{50}{\percent}$. \textit{Right}: Line profile of one of the solar lines. The $5$ points for which their contribution function is computed are indicated by big dots color-coded with the $T_{1/2}$ values.}
	\label{Fig:1}
\end{figure*}

\begin{table}[h!]
    \caption{Stellar parameters adopted for the spectral synthesis. The parameters are the effective temperature, $T_{\mathrm{eff}}$, surface gravity, $\log{g}$, metalicity, $[\mathrm{M}/\mathrm{H}]$, micro- and macroturbulences, $v_{\mathrm{mic}}$ and $v_{\mathrm{mac}}$, and projected rotational velocity, $v\sin{i}$.}
    \begin{threeparttable}
	\begin{tabular*}{\linewidth}{c @{\extracolsep{\fill}} c @{\extracolsep{\fill}} c}
		\toprule
		\midrule
		\textbf{Parameter}        & \textbf{Sun}  & \textbf{\aCenB{}} \\
		\midrule
		$T_{\mathrm{eff}}$ [K]    & 5770\tnote{a} & 5189\tnote{c}     \\
		$\log{g}$                 & 4.00\tnote{a} & 4.30\tnote{c}     \\
		$[\mathrm{M}/\mathrm{H}]$ & 0.00\tnote{a} & 0.22\tnote{c}     \\
		$v_{\mathrm{mic}}$ [km/s] & 0.85\tnote{b} & 0.95\tnote{c}     \\
		$v_{\mathrm{mac}}$ [km/s] & 3.98\tnote{b} & 4.87\tnote{b}     \\
		$v\sin{i}$ [km/s]         & 1.63\tnote{b} & 1.00\tnote{b}     \\
		\bottomrule
	\end{tabular*}
	\begin{tablenotes}
	\item[a] \footnotesize{\texttt{PySME} default.}
	\item[b] \footnotesize{\cite{Valenti&Fischer05}.}
	\item[c] \footnotesize{\cite{Morel18}.}
	\end{tablenotes}
	\end{threeparttable}
	\label{Tab:01}
\end{table}

\section{Methods}\label{Sect:3}

In this section, we describe how to compute, for each spectral point in the master spectrum, the stellar temperature below which $\SI{50}{\percent}$ of the emergent flux originates (see Sect.~\ref{Sect:3.1}), what we will hereafter call the average formation temperature. We thereafter investigate if the stellar activity signal depends on this average formation temperature by dividing each spectral line into a few average formation temperature bins, and measuring the RV for each of those bins. Finally, to reach the meter-per-second precision level in RV to be able to observe activity signals in the Sun and \aCenB{}, we combine the information of each average formation temperature bin obtained for all selected lines.

\subsection{Computing the average formation temperature}\label{Sect:3.1}

\begin{figure*}[t!]
	\includegraphics[width=\textwidth]{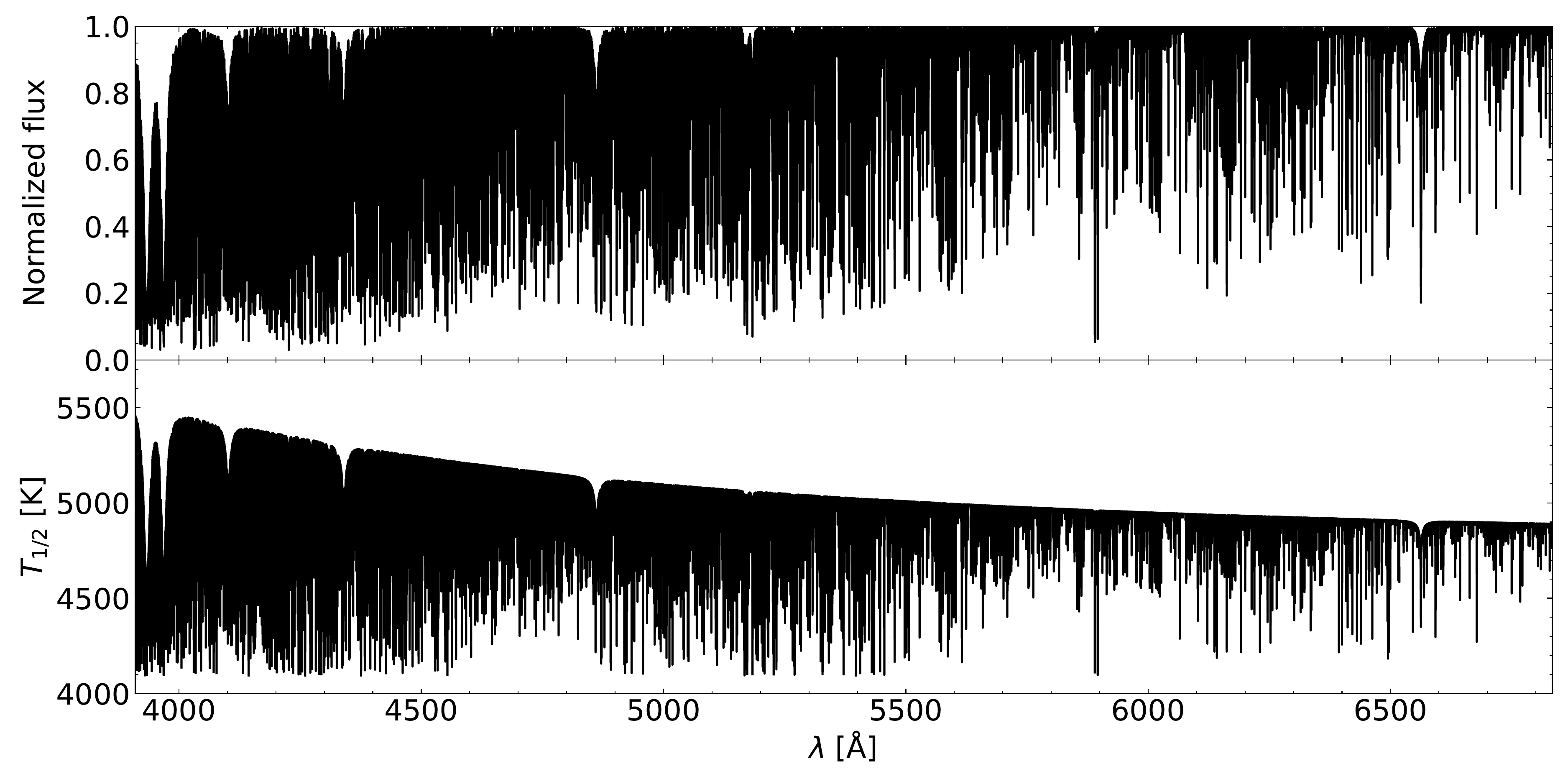}
	\caption{\textit{Top}: HARPS-N solar spectrum in units of normalized flux. \textit{Bottom}: Same spectrum as above converted to units of average formation temperature.}
	\label{Fig:2}
\end{figure*}

\texttt{PySME} automatically calculates the temperature profile, $T(\tau_0)$, on a grid of optical depths, $\tau$, where the zero subscript signifies a reference wavelength of $\SI{5000}{\angstrom}$. In order to evaluate other wavelengths, $\lambda$, the corresponding optical depth, $\tau_\lambda$, is given by \citep{Gray:2005},
\begin{equation}\label{Eq:1}
	\mathrm{d}\tau_\lambda = \frac{\kappa_\lambda}{\kappa_0}\mathrm{d}\tau_0 \, ,
\end{equation}
where $\kappa$ is the sum of the line and continuum opacities. The cumulative contribution function, $C(\tau_\lambda)$, which provides the emergent flux at a certain depth, is proportional to the source function, $S(\tau_\lambda)$, according to the following integral,
\begin{equation}\label{Eq:2}
	C(\tau_\lambda) \propto \int_{0}^{\tau_\lambda} S(\tau_\lambda')\mathrm{e}^{-\tau_\lambda'} \ \mathrm{d}\tau_\lambda' \, .
\end{equation}
Evaluating the source function, which is also provided by \texttt{PySME}, for a discrete grid in optical depths, the integral in Eq.~\ref{Eq:2} becomes a summation, and the contribution function can be normalized by dividing with its last point. Finally, by having both the temperature profile and contribution function computed on the same grid of optical depths for any given wavelength, a change of variable can be made to obtain the contribution function as a function of temperature. In essence, since we are implementing a 1D plane-parallel model atmosphere, both optical depth and temperature are equivalent variables in the sense that they are both monotonically increasing with geometrical depth (although at different rates). However, we chose to convert to temperature as a more intuitive metric which can also be easily related to the commonly known stellar effective temperatures. We refrain from attempting to convert to geometrical depth, as this involves the inclusion of wavelength-dependent stellar radii and their uncertainties.

As a diagnostic average formation, we extract the temperature at which the cumulative contribution function is equal to $\SI{50}{\percent}$, denoted $T_{1/2}$, which is obtained through a linear interpolation. Fig.~\ref{Fig:1} shows $T_{1/2}$ computed at $5$ different points across one of the solar lines, demonstrating that points close to the continuum are formed in hotter (deeper) parts of the photosphere, whereas the line core is formed in cooler (shallower) regions. Although Fig.~\ref{Fig:1} only demonstrates the derived $T_{1/2}$ value for $5$ line points across a single spectral line, the calculations are in fact repeated for every single observed wavelength point of the HARPS and HARPS-N spectra. The bottom panel of Fig.~\ref{Fig:2} shows the entire HARPS-N solar spectrum converted into units of $T_{1/2}$ rather than flux. It can clearly be seen that the formation temperature roughly traces the shapes of the spectral lines locally. The global decrease at longer wavelengths is due to the increase of the continuum opacity, which in the optical wavelength range is dominated by the bound-free absorption of the negative hydrogen ion \citep{Gray:2005}.

\subsection{Line selection}\label{Sect:3.2}

\begin{figure*}[t!]
	\includegraphics[width=\textwidth]{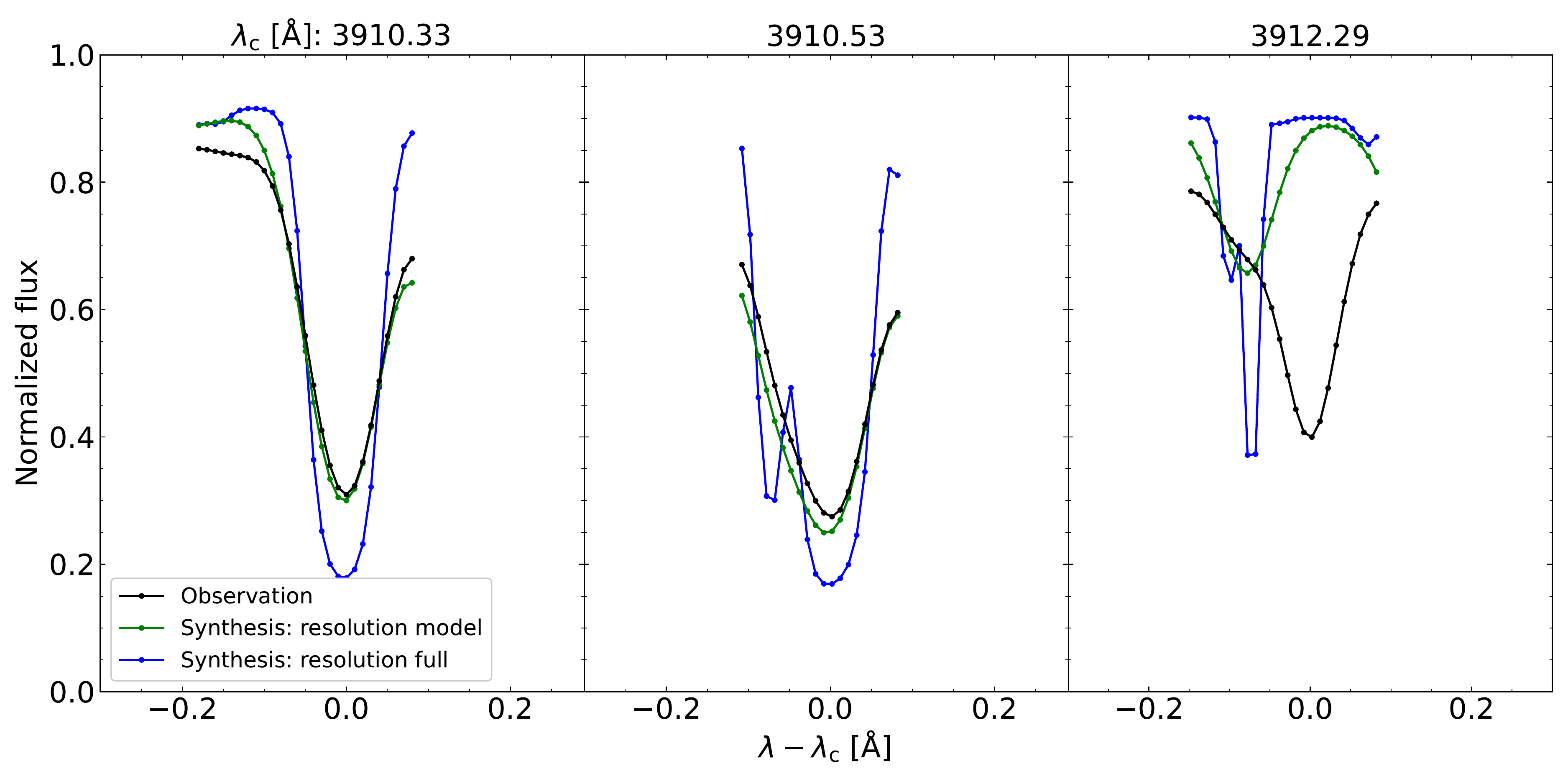}
	\caption{Observed (black) and synthesized (green and blue) spectra shown at the position of three exemplary spectral lines. The synthesis including micro- and macroturbulences, rotational broadening and generated at the instrumental resolution of HARPS and HARPS-N is shown in green. The synthesis without turbulence and rotational broadening, and at the highest resolution for the observational wavelength grid is shown in blue. \textit{Left}: Line considered unblended and included in our study. \textit{Middle}: Line rejected due to a blend seen in the unbroadened synthesis. \textit{Right}: Line rejected due to large discrepancy between the observation and the broadened synthesis.}
	\label{Fig:3}
\end{figure*}

\begin{figure*}[t!]
	\includegraphics[width=\textwidth]{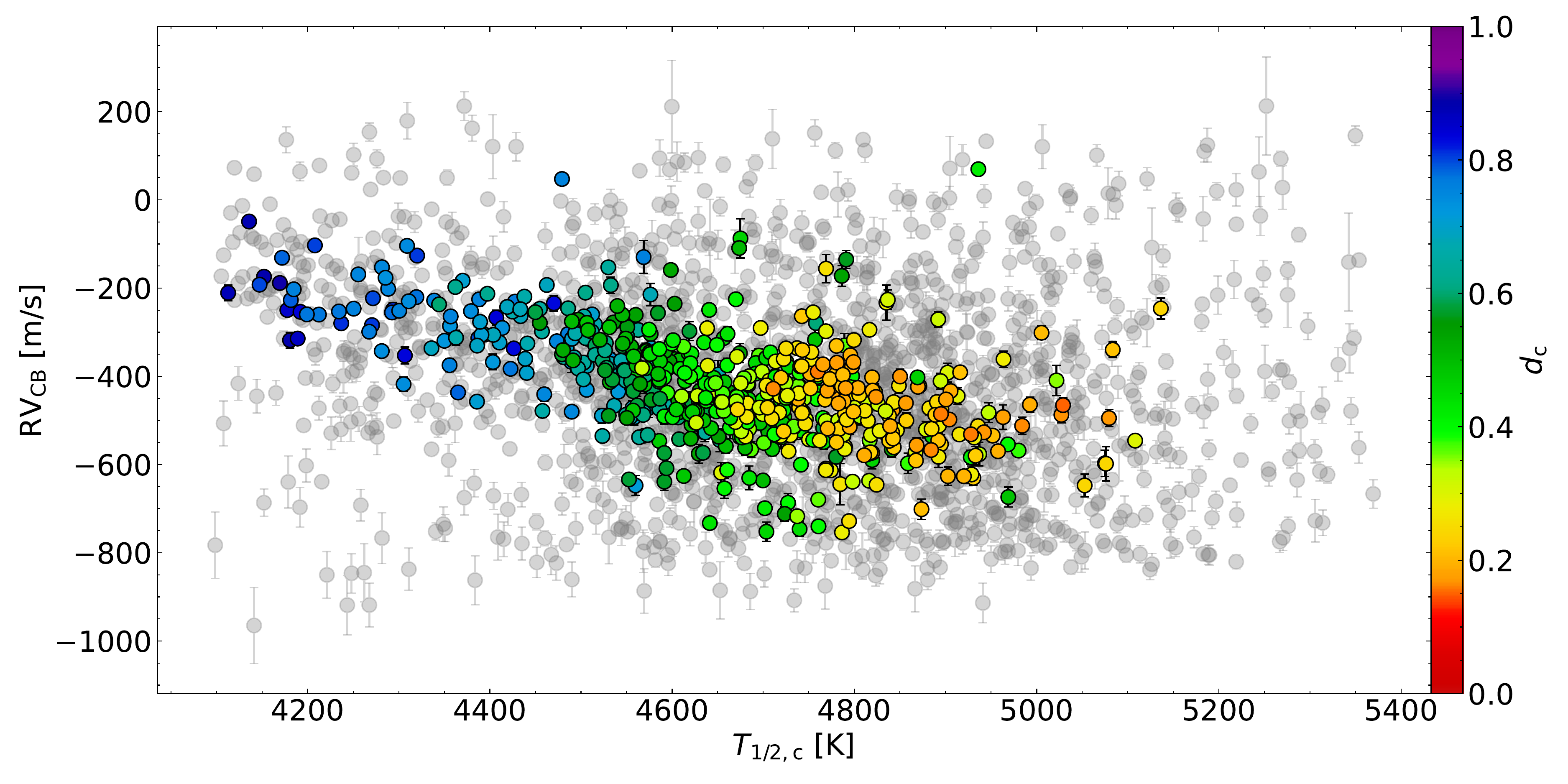}
	\caption{Solar convective blueshift as a function of $T_{1/2}$ at the line cores. Only lines which could be cross-matched with the VALD3 line list are displayed. The markers of lines from the final selection are color-coded with the line depth shown in the colorbar, the other lines are shown in gray. The gravitational redshift of $\mathrm{RV}_{\mathrm{grav}}{=}GM/Rc$, where $M$ and $R$ are the solar mass and radius, has been subtracted.}
	\label{Fig:4}
\end{figure*}

\begin{figure*}[t!]
	\includegraphics[width=\textwidth]{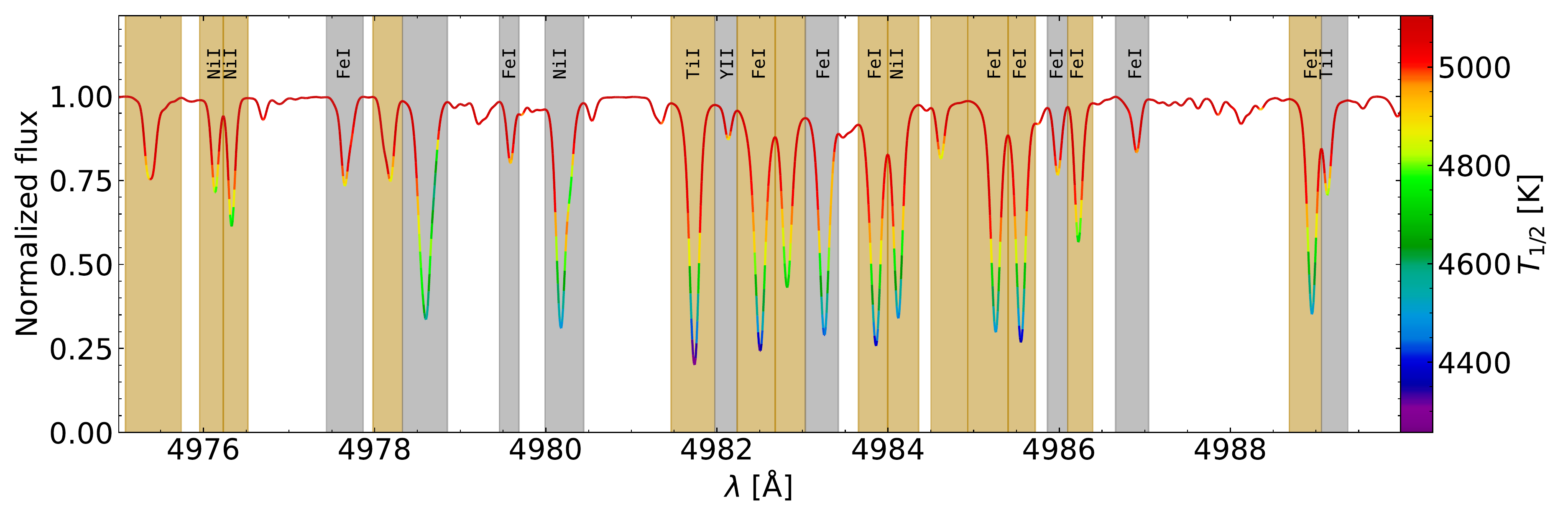}
	\caption{$\SI{15}{\angstrom}$ spectral window color-coded according to the average formation temperature. The shaded areas indicate whether a spectral line is selected (beige) or rejected (grey). The labels above some spectral lines specify the element and ionization of identified lines.}
	\label{Fig:5}
\end{figure*}

In order to properly asses the average formation temperatures for each spectral point, we need to consider lines that are well modeled by the spectral synthesis. To identify spectral lines that are unblended and symmetric, we initially used the morphological parametrisation described in Appendix C of \cite{Cretignier+20a}. Starting from this line selection, we then consider our two syntheses. For each spectral line, we check if the high-resolution profile contains multiple local minima, in which case we reject the line due to blending. Additionally, we also measure the line-depth weighed $\chi^2$ between the instrumental-resolution profile and the observed profile, and reject all the lines for which the $\chi^2$ is more than $4\sigma$ larger than the median value, which can occur due to faulty atomic parameters in the VALD3 database. Fig.~\ref{Fig:3} demonstrates the comparison between observed and synthesized solar spectra for an accepted line, a line rejected because of hidden blends and a line rejected because of large discrepancy.

For each spectral line in the master spectrum, the CB is computed by fitting a second-order polynomial to the center-most $7$ points. The CB is derived from the difference between the fitted center and the laboratory wavelengths found from a cross-match with the VALD3 line list. 

Fig.~\ref{Fig:4} shows the solar CB versus $T_{1/2}$ interpolated at the line cores, with errors estimated from the uncertainties of the polynomial coefficients. The figure showcases how our selection of unblended lines clearly accentuates the dependency of CB on certain line properties, as well as how the relation is transformed from non-linear to linear when plotted against formation temperature as opposed to the more common line depth \citep[see e.g. Fig. 3 in][]{Reiners+16}. The CB of \aCenB{}, shown in Fig.~\ref{Fig:B1}, follows the same relation, although with a smaller slope than for the Sun which reflects the expected behaviour of more extensive convective envelopes and thus smaller convective velocity gradients in stars with a later spectral type \citep[e.g.][]{Meunier+17b,Liebing+21}. The CB of \aCenB{} does not, however, reach near-zero values at low temperatures, which is likely due to the difficulty of constraining the gravitational redshift and orbital motion of the two main components in the Alpha Centauri system \citep[e.g.][]{Pourbaix+02}.

Although the linearity with temperature appears clear when only considering lines for which the estimated core formation temperature is assumed to be reliable, we are alike previous studies \citep[e.g.][]{Gray09,Reiners+16} unable to provide a physical explanation for its origin. To verify the interpretation of the trend, namely that the convective velocity increases linearly with the stellar temperature profile, one could suggestively compare with more complex, magneto-hydrodynamical (MHD) stellar models which take convection into consideration. Such a model comparison is outside the scope of this study, and we note that the following analysis is ultimately unaffected by the CB since our RV time series are derived with respect to a master spectrum, for which the absolute shift is of lesser importance. The CB, which provides the shift relative to laboratory measurements, is simply presented here to validate our line selection and derivation of average formation temperature. By showing its inverse proportionality with respect to the core formation temperature, which on average increases with decreasing line depth, we are able to demonstrate consistency with the literature.

\subsection{RV per temperature bin for individual spectral lines}\label{Sect:3.3}

Once $T_{1/2}$ is calculated for all points of the master spectrum (see Sect.~\ref{Sect:3.1}), our goal is to measure the RV for different bins in average formation temperature as a function of time, by measuring it for each individual spectrum from which the master is built. The RV is derived by template matching as described in \citet{Dumusque18} and \citet{Cretignier+20a}, using the master spectrum as the template. We construct the temperature bins by heuristically dividing the entire range of $T_{1/2}$ values (for the entire spectrum) into $N$ intervals of equal length, where $N$ is a positive integer. Then, for each line profile we measure the RVs of separate segments, by identifying which points are formed within the unique temperature bins. We remark that each line will at most be divided into $N$ segments, however, the majority of lines will only have points in a subset of bins. Generally, the core will constitute the line segment formed at the coolest temperature bin, whereas hotter temperature bins will have nearly symmetrical contributions from the left and right wings. As an example, if we assume the line in Fig.~\ref{Fig:1} is divided into $3$ segments, then points near dot $3$ will be associated to a low temperature bin, points near dots $2$ and $4$ will be associated to a medium temperature bin, and points nears dots $1$ and $5$ will be associated to a high temperature bin. Line segments with less than three points are discarded, to have a good estimate of the flux derivative needed to compute the RVs \citep[see][]{Bouchy+01}. The end product is an RV matrix with dimensions $N_{\mathrm{spec}}{\times}N_{\mathrm{line}}{\times}N_{\mathrm{temp}}$, where $N_{\mathrm{spec}}$ is the number of spectra in the time series, $N_{\mathrm{line}}$ is the number of spectral lines, and $N_{\mathrm{temp}}$ is the selected number of temperature bins, for which some of the entries are empty due to a lack of line points. A matrix of equal size is created for the corresponding RV uncertainties. Averaging the RV measurement of all lines for a given temperature bin (see Sect.~\ref{Sect:3.4}) allows us to reach high RV precision, and therefore to investigate if the stellar activity signal has a dependence on average formation temperature.

\begin{figure*}[t!]
	\includegraphics[width=\textwidth]{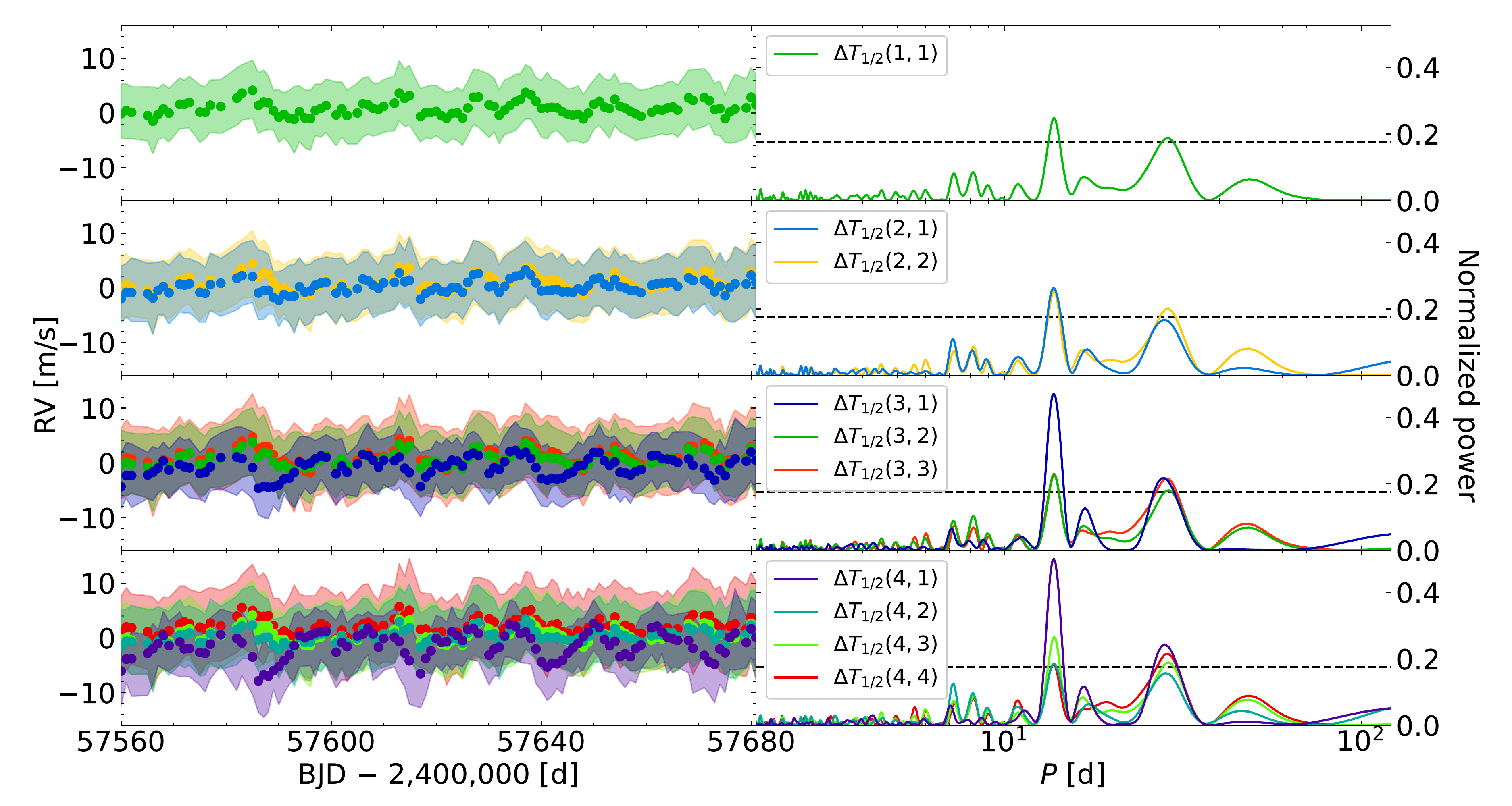}
	\caption{\textit{Left}: Sun RV time series for $1$ to $4$ temperature bins. Each point is the weighted average of all spectral lines with RV values for a given bin. The shaded intervals represent the $16\textsuperscript{th}$ to $84\textsuperscript{th}$ percentile of all RVs at any given time point. Times are given in \acrfull*{BJD}. \textit{Right}: GLS periodograms of the time series shown in the left panels. The legend displays the temperature bin symbol (see main text and Table~\ref{Tab:02} for definition). The dashed line indicates the $\SI{1}{\percent}$ FAP level.}
	\label{Fig:6}
\end{figure*}

\begin{table*}[t!]
    \caption{Specifications of the temperature-binned RV time series in Fig.~\ref{Fig:6}. For each configuration, the columns specify the number of temperature bins, the bin symbols as described in the text, the intervals in $T_{1/2}$, the total number of lines considered for all bins, the fractions of lines considered for each bin individually, and the RV RMS for the chosen time interval.}
	\begin{tabular*}{\textwidth}{c @{\extracolsep{\fill}} c @{\extracolsep{\fill}} c @{\extracolsep{\fill}} c @{\extracolsep{\fill}} c @{\extracolsep{\fill}} c}
		\toprule
		\midrule
		\textbf{Nr. of bins} & \textbf{Symbol} & \textbf{$T_{1/2}$ [K]} & \textbf{Nr. of lines} & \textbf{Fraction of lines [\%]} & \textbf{RMS [m/s]}\\
		\midrule
		\multirow{1}{*}{1} & $\Delta T_{1/2}(1,1)$ & 4095-5457 & \multirow{1}{*}{773} & 100 & 1.25\\
		\midrule
		\multirow{2}{*}{2} & $\Delta T_{1/2}(2,1)$ & 4095-4776 & \multirow{2}{*}{771} &  55 & 1.17\\
		                   & $\Delta T_{1/2}(2,2)$ & 4776-5457 &                      & 100 & 1.32\\
		\midrule
		\multirow{3}{*}{3} & $\Delta T_{1/2}(3,1)$ & 4095-4549 & \multirow{3}{*}{781} &  16 & 1.69\\
		                   & $\Delta T_{1/2}(3,2)$ & 4549-5003 &                      &  86 & 1.22\\
		                   & $\Delta T_{1/2}(3,3)$ & 5003-5457 &                      &  67 & 1.39\\
		\midrule
		\multirow{4}{*}{4} & $\Delta T_{1/2}(4,1)$ & 4095-4435 & \multirow{4}{*}{780} &   8 & 2.33\\
		                   & $\Delta T_{1/2}(4,2)$ & 4435-4776 &                      &  54 & 1.13\\
		                   & $\Delta T_{1/2}(4,3)$ & 4776-5116 &                      &  79 & 1.32\\
		                   & $\Delta T_{1/2}(4,4)$ & 5116-5457 &                      &  37 & 1.41\\
		\bottomrule
	\end{tabular*}
	\label{Tab:02}
\end{table*}

Identifying line segments on the master spectrum and then using those for each individual spectrum, could raise concerns about different spectral sampling and Doppler shifts between the spectra. The first problem is not an issue as the master spectrum and the individual spectra are sampled on the same wavelength grid. Regarding the shifts, although the master spectrum and the individual spectra are all in the same restframe, stellar activity will induce at most, in the case of our target stars, peak-to-peak Doppler shifts of a few dozens of meters-per-seconds. Such shifts represent less than a tenth of a pixel (${\sim}\SI{820}{\meter/\second}$ for HARPS and HARPS-N), and will thus not inflict with the identification of line segments.

Although line profiles in units of average formation temperature closely resemble their flux counterparts, the idea behind using temperature bins rather than flux bins is that the two quantities are inherently different both locally for single lines and globally for the entire spectrum. If one were to compare two lines of equal line depths yet unequal atomic properties (such as elemental species, ionization stage, excitation energies, among others), they would not necessarily be formed at the same temperature, which depends on the line opacity and hence on these individual properties. This can be seen in Fig.~\ref{Fig:5}, where the flux profiles of a few lines have been color-coded with their $T_{1/2}$ values. The cores of shallow lines are not colored the same as the wings of deeper lines at equal flux levels. Likewise, as was shown in Fig.~\ref{Fig:2}, when considering the entire spectrum, there are also wavelength-dependent variations of the continuum opacity. Both of these effects motive the usage of formation temperature as a preferred variable for probing photospheric height, and the types of activity occurring at various heights.

\subsection{RV per temperature bin for spectral line average}\label{Sect:3.4}

Before averaging line segments from the same formation temperature bin together to increase the RV precision, we curate the selection of line segments included in the averaging. For each RV time series per line and per temperature bin, we remove outliers in the RVs and their uncertainties through an iterative $4\sigma$ clipping (the clipping is only upper-ended for the uncertainties, i.e. we only remove points for which the uncertainty is $4$ standard deviations larger than the median errorbar). RV time series which contain more than $\SI{5}{\percent}$ outliers are entirely rejected. In addition, time series with a $z$-score, defined as the RV \acrfull*{RMS} divided by the median uncertainty, smaller than $2$ are also excluded before averaging to rejected time series with no significant information. We note that for the $4$ bin configuration, the $\SI{5}{\percent}$ outlier criterion rejects an average $\SI{5.5}{\percent}$ of RV time series, while the $z$-score criterion rejects $\SI{9.9}{\percent}$ of them.

The average RV time series per temperature bin is then obtained by averaging the RVs of all remaining spectral line segments, weighting by their inverse variance at each time point. This results in an RV matrix with dimensions $N_{\mathrm{spec}}{\times}N_{\mathrm{temp}}$.

\section{Results}\label{Sect:4}

\subsection{Short-term RVs of the Sun}\label{Sect:4.1}

Short-term RV measurements of the Sun are performed on $105$ daily-binned spectra taken between Jun. 20 and Oct. 19, 2016, in order to cover a few solar rotational periods during an active solar phase. Fig.~\ref{Fig:6} shows the inverse-variance weighted RV time series of all considered lines for when the formation temperature is increasingly divided into $1$ to $4$ bins of equal length (see Sect.~\ref{Sect:3.3}), denoted $\Delta T_{1/2}(N,n)$, where $N$ is the number of bins and $1 {\leq} n {\leq} N$ is the bin index. The figure also shows the corresponding normalized, \acrlong*{GLS} \citep[GLS;][]{Zechmeister&Kurster09} periodograms for each time series with the $\SI{1}{\percent}$ \acrfull*{FAP} indicated. The uppermost left panel of Fig.~\ref{Fig:6}, for the configuration of just $1$ temperature bin (equivalent to no line segmentation based on formation temperature), is thus showing the same RV time series as other traditional methods of measurements would produce, i.e. by measuring LBL RVs and averaging over all lines, or by cross-correlating with a spectral type mask. The periodogram in the uppermost right panel is also showing two expected peaks at the solar synodic rotation period of ${\sim}\SI{27}{d}$ and half of it (the first harmonic). As the RVs are separated into an increasingly larger number of temperature bins (second, third and fourth rows in Fig.~\ref{Fig:6}), we note the emergence of a signal in the coolest temperature bin which is anti-correlated with respect to the total signal and has a significantly stronger periodogram peak at half the rotation period.

Table~\ref{Tab:02} lists the temperature intervals used for each bin configuration, the number of selected lines (which differs between configurations depending on the sigma-clipping and $z$-score criteria described above), and the RV RMS for each bin. For the configuration with $4$ bins, the RMS verifies that the coolest and hottest bins, ${\Delta}T_{1/2}(4,1)$ and ${\Delta}T_{1/2}(4,4)$, show larger dispersion compared to the intermediate bins. Even though the bin averages are calculated using varying number of lines (due to some bins only existing for a certain subset of lines), the $z$-score criteria ensures that larger uncertainties are only included when the RMS is up-scaled by at least the same amount.

While Fig.~\ref{Fig:6} provides a visualisation of the collective (or rather, average) behaviour of all considered lines over time, it does not inform us about the individual lines which contribute to each bin. Fig.~\ref{Fig:A1} shows a corner plot of various line parameters where each dot is an individual line. The dots are colored according to the temperature bin in which they belong, for the configuration with $4$ bins (same as the bottom panel of Fig.~\ref{Fig:6}). Regarding the distribution of the various line parameters and their co-variances, we state the following remarks:
\begin{itemize}
    \item Central wavelength, $\lambda_{\mathrm{c}}$: All bins except the hottest (red) does not have a dependency on chromaticity. This can be explained by the gradient of the continuum opacity (see Fig.~\ref{Fig:2}) causing the hottest bin to be biased toward shorter wavelengths.
    \item Line depth, $d_{\mathrm{c}}$: All bins include lines of any line depth, except the coolest (purple) bin which primarily consists of the cores of the strongest lines.
    \item Convective blueshift, $\mathrm{RV}_{\mathrm{CB}}$: Increases (i.e. approaches zero due to its negative nature) gradually with decreasing temperature. The line segments from the coolest bin are the least affected by CB.
    \item RV RMS for the time interval in Fig.~\ref{Fig:6}: Similar distributions for all bins. Line segments with the smallest scatter originate from an intermediate (cyan) bin.
    \item Pearson correlation between LBL RV and mean RV (i.e. average total RV without temperature binning, same as the upper panel of Fig.~\ref{Fig:6}), $\mathcal{R}(\mathrm{RV}_i, \langle\mathrm{RV}\rangle)$: Most lines segments are weakly correlated with the average RV, except those from the coolest bin which are evenly spread around zero correlation.
    \item Pearson correlation between LBL RV and \ion{Ca}{II} H\&K $S$-index, $\mathcal{R}(\mathrm{RV}_i, S)$: Most line segments are weakly correlated with the $S$-index, except those from the coolest bin which are mostly weakly anti-correlated.
\end{itemize}

\subsection{Solar simulations with \texttt{SOAP-GPU}}\label{Sect:4.2}

\begin{figure}[t!]
	\includegraphics[width=\linewidth]{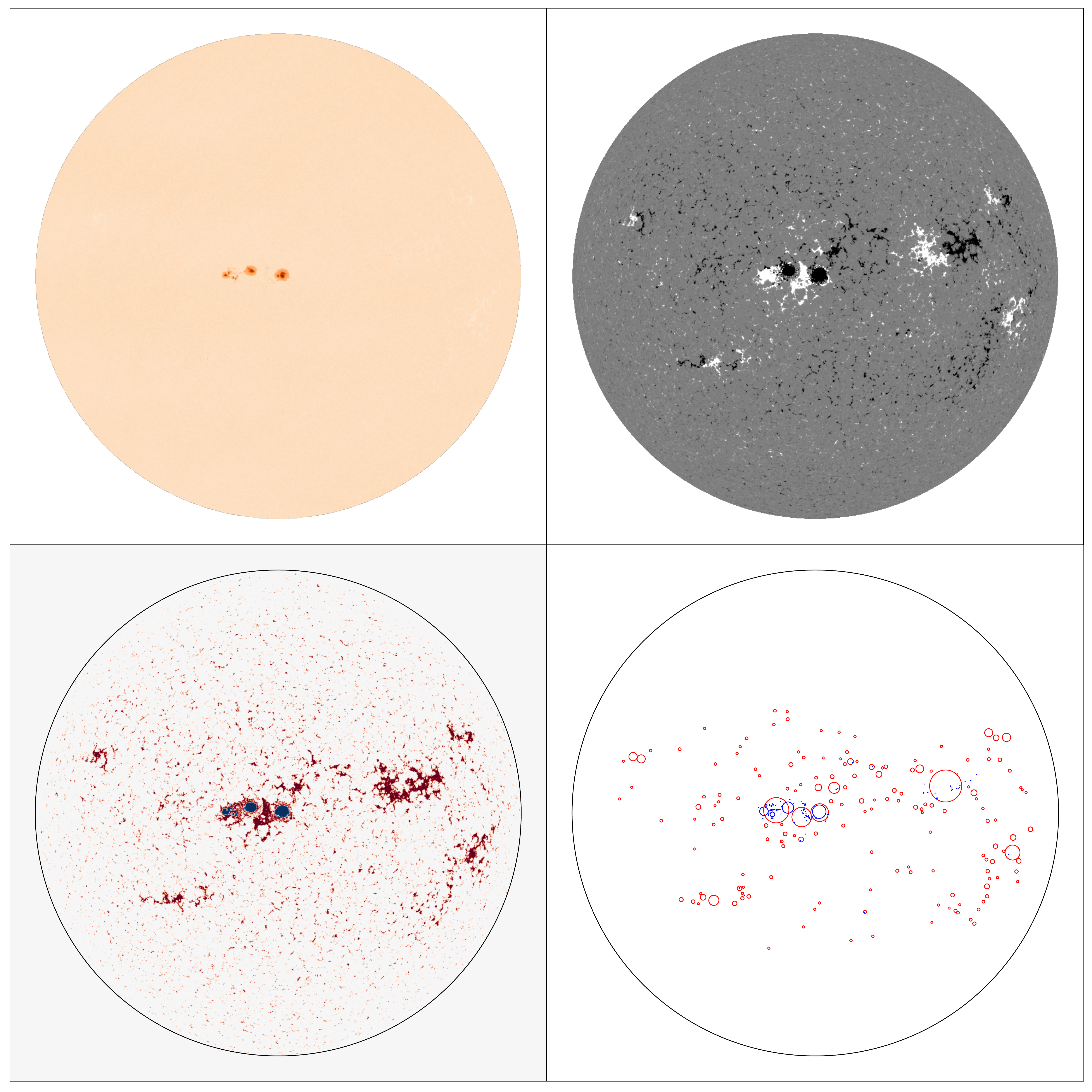}
	\caption{\textit{Top panels}: Flattened intensitygram (left) and magnetogram (right) of the Sun taken by SDO/HMI on Jul. 17, 2016 at UT12:00:00. \textit{Bottom left panel}: Pixels identified as spots (blue) or faculae (red). \textit{Bottom right panel}: Circles representing the same area for spots (blue) and large faculae (red) as the panel to the left, used as \texttt{SOAP-GPU} input.}
	\label{Fig:7}
\end{figure}

To enable the comparison between the temperature-binned RVs and the impact of different active regions, we implement solar simulations for the same time interval as in Sect.~\ref{Sect:4.1} with the \texttt{SOAP-GPU} code \citep{Zhao&DumusquePrep}, a GPU-based successor to the \texttt{SOAP 2.0} code \citep[\acrlong*{SOAP},][]{Dumusque+14}. \texttt{SOAP-GPU} simulates RVs by modeling the observed solar disk as a grid of pixels. Each pixel is assigned a line-of-sight velocity from the combined contributions of rotation and CB. The code then injects each pixel with one of two spatially resolved solar spectra, either a quiet Sun spectrum or a sunspot spectrum, which are Doppler shifted according to the velocity of each pixel. A disk-integrated spectrum is computed by summing all pixels, after scaling the intensity levels with limb darkening and lower/higher contrast for the pixels selected as spot/facula pixels.

From the \acrfull*{HMI} instrument onboard the \acrfull*{SDO}\footnote{Available at \url{https://sdo.gsfc.nasa.gov}.}, we retrieve continuum intensitygrams and magnetograms with a cadence of $\SI{720}{\second}$. We retrieve one frame at noon for every day available between the dates used in Sect.~\ref{Sect:4.1}. We then follow the methodology in \cite{Haywood+16} to extract a threshold map of spot locations from the flattened intensitygrams and facula locations from the unsigned magnetograms. Spot pixels are identified as those where the intensity is below $\SI{89}{\percent}$ of the quiet Sun intensity, and faculae pixels as those where the unsigned magnetic field is larger than $3$ times its standard deviation. With a clustering algorithm, we search for pixel clusters in the threshold maps to find the most prominent regions in each set of images. Isolated pixels are rejected due to possible fluctuations. Large faculae are defined as regions covering at least $\SI{20}{\micro Hem}$, the rest of the selected pixels being associated to the network covering almost the entire disk. For each spot and large facula, the spherical coordinates and radius of the corresponding circle with equal area (see Fig.~\ref{Fig:7}) are computed and given as the sole input to \texttt{SOAP-GPU}.

\begin{figure}[t!]
	\includegraphics[width=\linewidth]{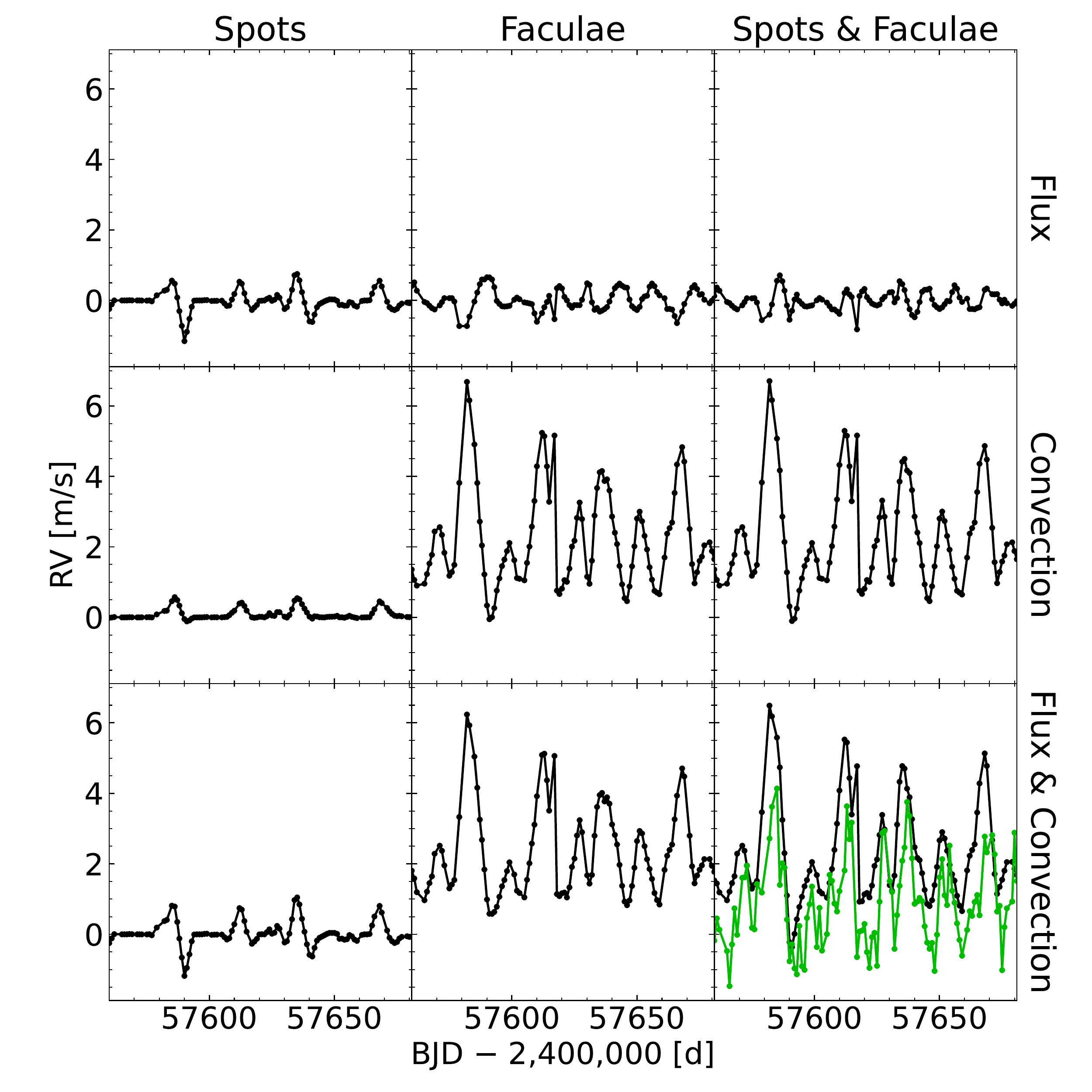}
	\caption{\texttt{SOAP-GPU} generated RVs for the same time points as in Fig.~\ref{Fig:6}, separated into active regions (spots, faculae and both) and effects (flux, convection and both). The green curve in the lower right panel is the observed RVs without temperature binning.}
	\label{Fig:8}
\end{figure}

\begin{figure}[t!]
	\includegraphics[width=\linewidth]{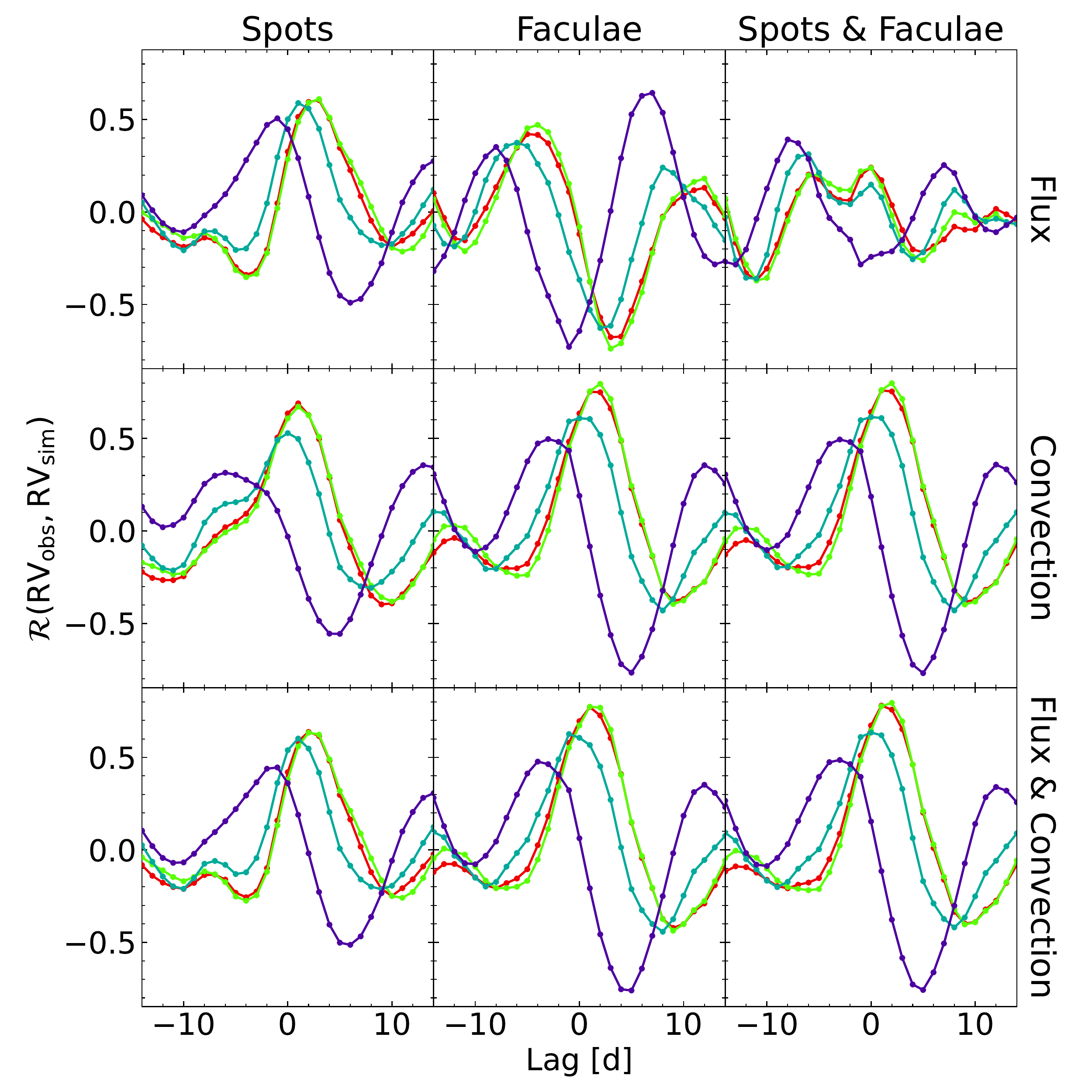}
	\caption{Pearson correlation between observed and simulated RVs as a function of time lag. The $4$ temperature bins are represented by the same colors as in Fig.~\ref{Fig:6}.}
	\label{Fig:9}
\end{figure}

\texttt{SOAP-GPU} enables the total RV signal to be separated into the contributions of different active regions (i.e. spots, faculae or both combined) and effects (i.e. flux, convection or both combined), shown in Fig.~\ref{Fig:8}. We thereafter linearly interpolate the simulated RVs onto the same time sampling as the observations and cross-correlate the temperature-binned RV time series with the simulated contribution of every combination of region/effect by computing the Pearson correlation coefficient for each configuration while allowing for a phase lag going from $-14$ to $14$ days (covering just over one rotational period) in steps of $1$ day (see Fig.~\ref{Fig:9}). We find that the two hottest temperature bins are strongly correlated with the convective effect of faculae, the second-to-coolest bin to not be strongly correlated with any type of contribution, and the coolest bin to be ambiguously anti-correlated with faculae flux with a lag of $-1$ day or faculae convective inhibition with a lag of $5$ days. It is intuitively difficult to argue why stellar activity signals would propagate from one temperature region to the next on a timescale of several days, therefore, it is more likely that the coolest bin in temperature is related to the flux effect induced by faculae.

\subsection{Long-term RVs of the Sun}\label{Sect:4.3}

\begin{figure*}[t!]
	\includegraphics[width=0.49\textwidth]{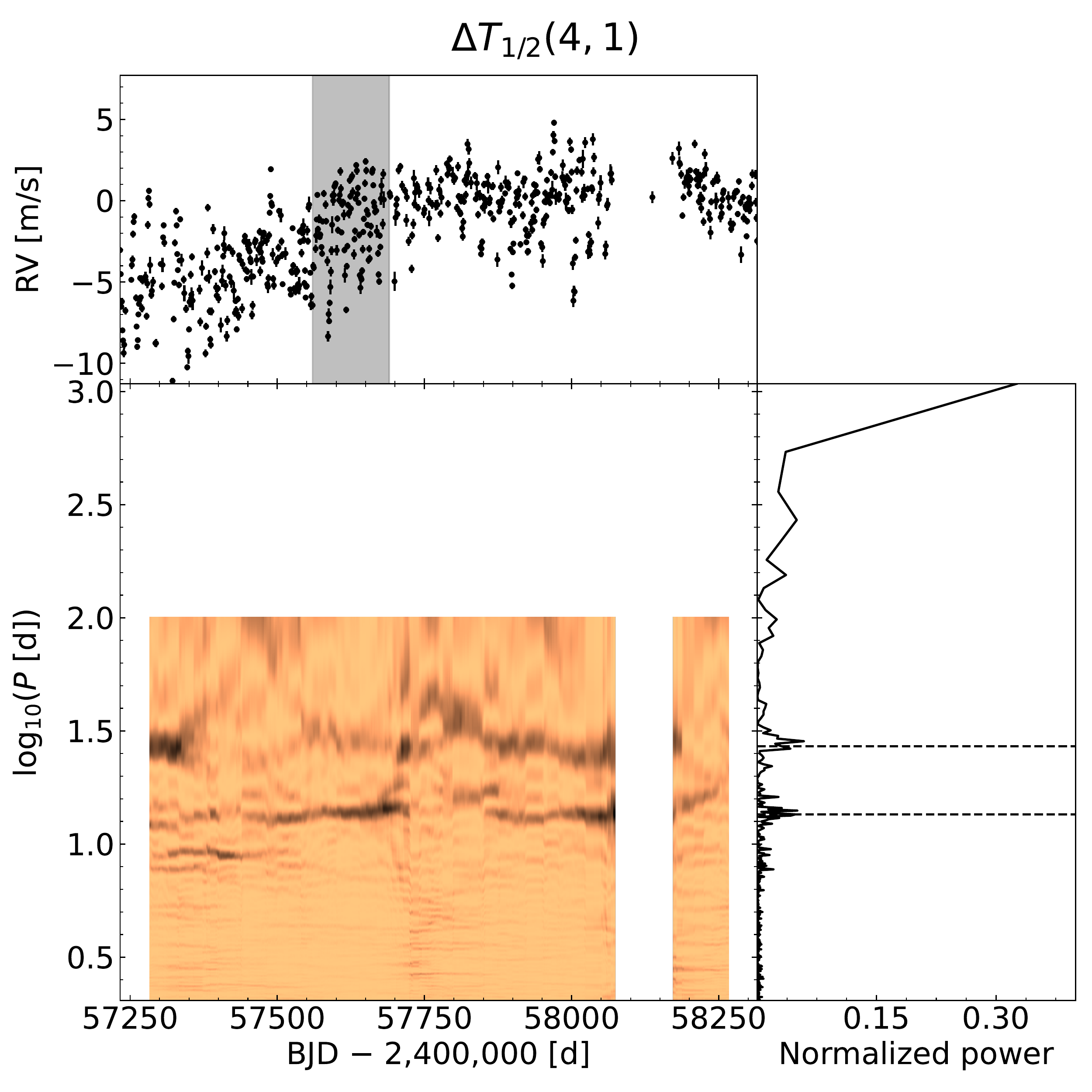}
	\includegraphics[width=0.49\textwidth]{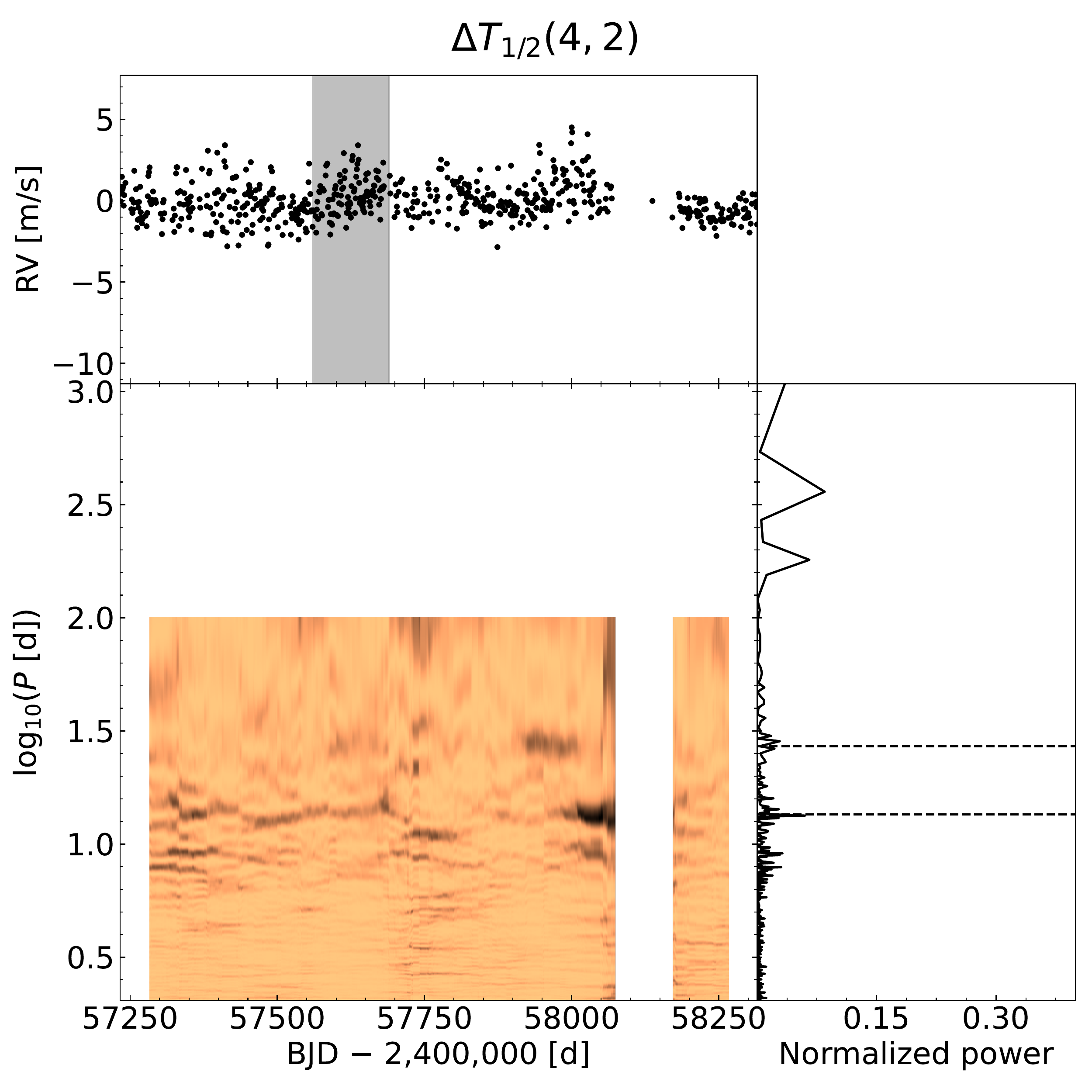}
	\includegraphics[width=0.49\textwidth]{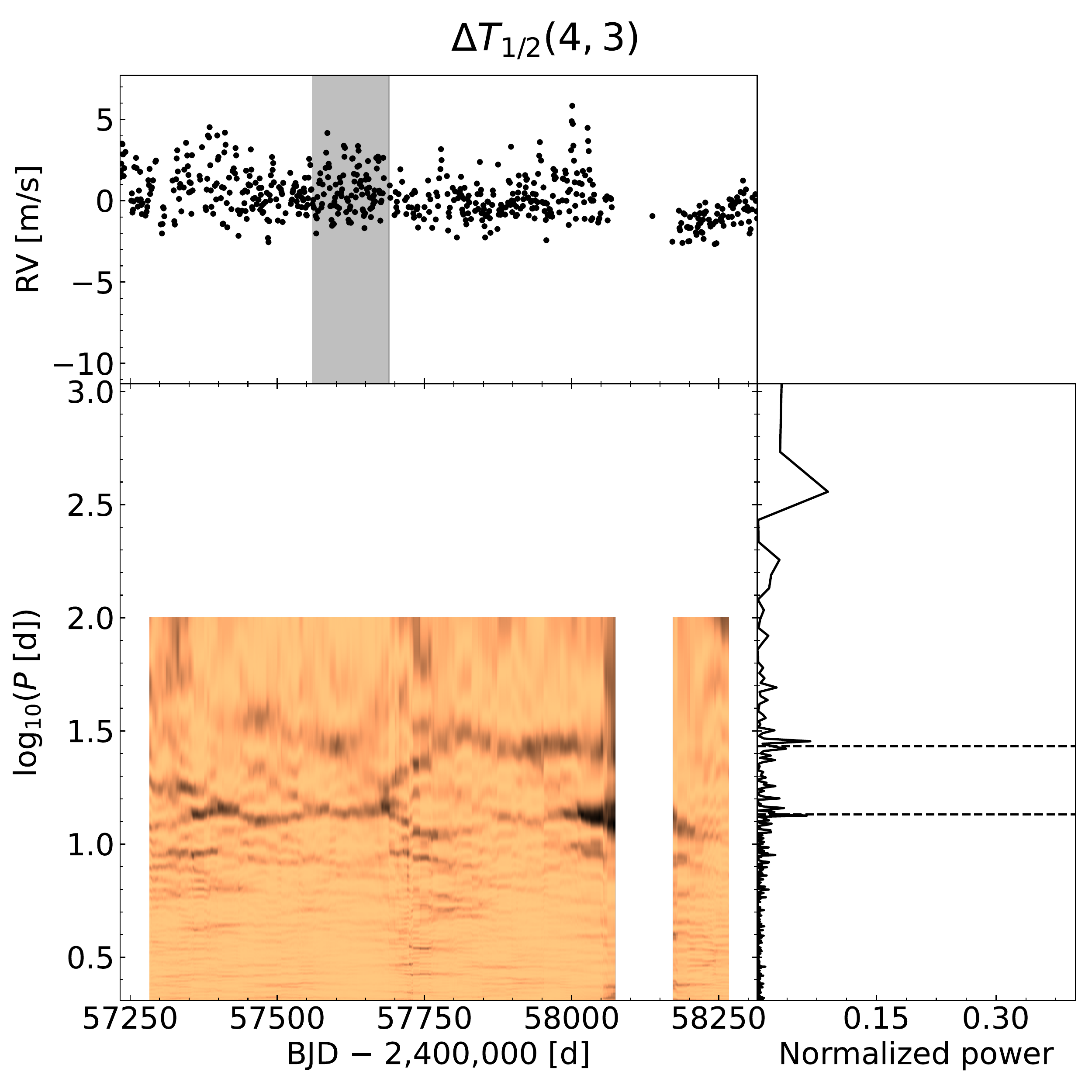}
	\includegraphics[width=0.49\textwidth]{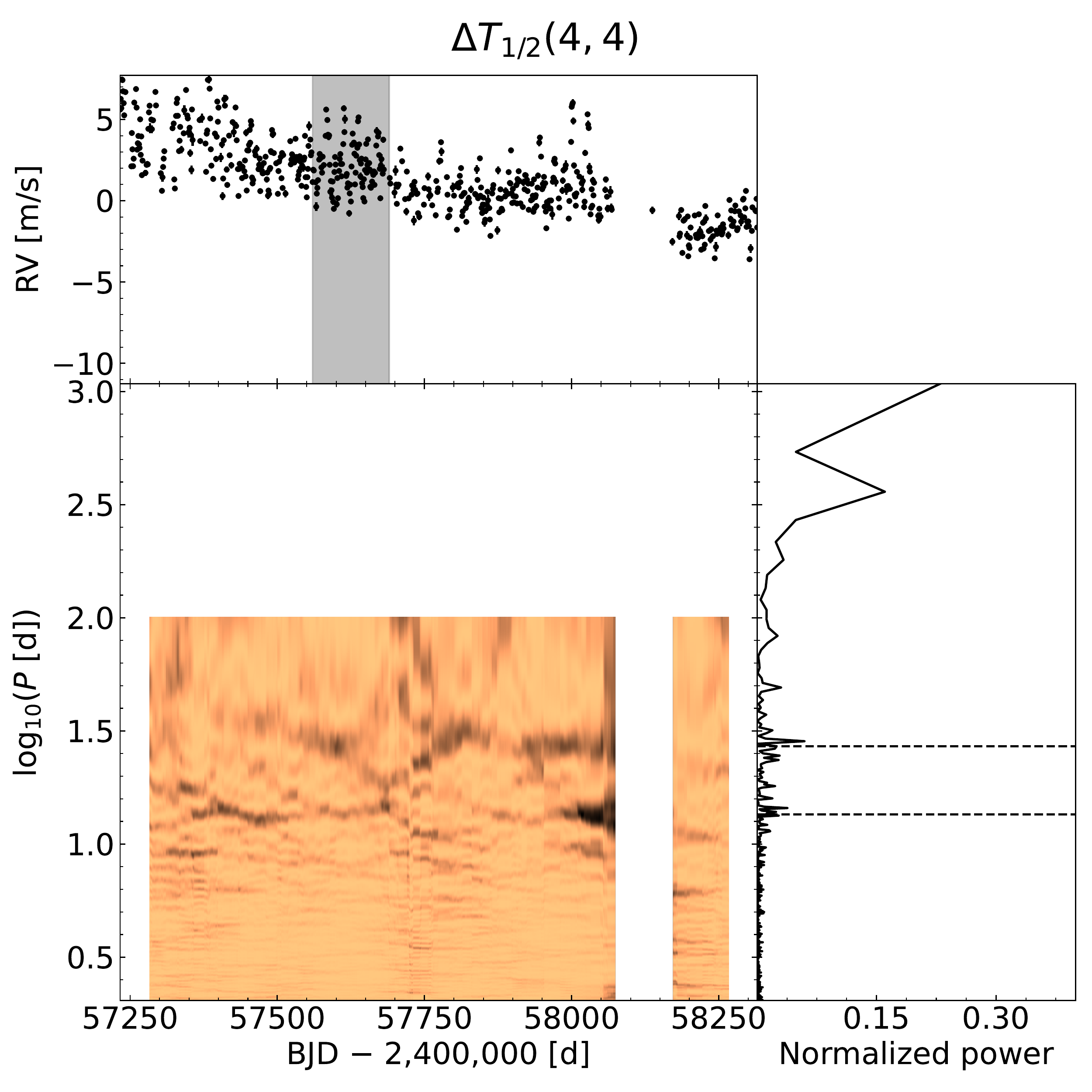}
	\caption{Sliding periodogram of the temperature-binned RV time series of the Sun. The upper panels show the total RV time series with the short-term interval used in Sect.~\ref{Sect:4.1} indicated by the gray, shaded area. The right panels show the GLS periodogram of the total time series, with the solar rotation period and its first harmonic indicated by the dashed lines. The columns of the central colormaps show the GLS periodogram of every $100$-day window, where darker colors indicate stronger power. The window is shifted by steps of 1 day, and gaps correspond to windows with more than $\SI{25}{\percent}$ of the data points missing.}
	\label{Fig:10}
\end{figure*}

Long-term RV measurements of the Sun are performed on $607$ daily-binned spectra taken between Jul. 29, 2015 and Jul. 15, 2018, spanning almost $3$ years near the end of Solar Cycle 24. In order to simultaneously visualize the variability which was observed on the short-term timescale in Sect.~\ref{Sect:4.1} over several similar intervals, and the overall variability when considering the total available time span, we make use of a sliding periodogram, as shown in Fig.~\ref{Fig:10} for the configuration of $4$ temperature bins. The sliding periodogram consists primarily of a color map, where each column represents the GLS periodogram of a fixed time window (here set to $100$ days to equate a few solar rotation periods alike the short-term interval). The window is slid $1$ day for each column, and windows containing less than $\SI{75}{\percent}$ of the expected number of data points (i.e. $75$ points for our window size of $100$ days) are left as blank columns. The color map will thus always have gaps of $\SI{25}{\percent}$ of the window size on the left- and right-most edges, and its vertical extent (the maximum fitted period) will be equal to the window size itself. The upper panel of the periodogram shows the entire RV time series with its conventional periodogram displayed in the right panel.

The sliding periodograms indicate that the $100$-day windows show consistent power distribution over approximately the first half of the covered time period, corresponding to the near end of the most recent solar cycle during which the Sun showed significant levels of activity and active region coverage. On the overall time span, we find the most extreme temperature bins (the coolest and hottest) to be dominated by the long-term magnetic cycle, and apparently oppositely affected similar to their short-term behaviors. The intermediate bins are less affected by low-frequency signals while still showing relatively strong amplitudes at half the solar rotation period.

\subsection{RVs of \aCenB{}}\label{Sect:4.4}

RV measurements of \aCenB{} are performed on the 2010 observations which have been extensively studied \citep[see e.g.][]{Dumusque18,Cretignier+20a}. The RV curation of \aCenB{} follows the same steps as for the Sun, with the results shown in Appendix~\ref{Sect:B}.

We find a similar behaviour for the temperature-binned RVs, with the coolest bin being even more strongly anti-correlated with the $S$-index. The main difference for \aCenB{}, however, is the lack of power in the periodograms at around half the rotation period. This could be explained by the stellar inclination of \aCenB{}, believed to be $\SI{45}{\degree}^{+9}_{-19}$ \citep{Dumusque14}, weakening the impact of the flux effect due to a decreased rotational velocity component along the line of sight. Such an inclination also causes active regions near the northern pole to remain visible during the entire rotation, therefore amplifying the signal at the rotation period, and not half of it like for the equator-on Sun.

\section{Discussion}\label{Sect:5}

The analysis performed in this paper implies that the activity-induced, temporal RV signal is dependent on the formation temperature inside the stellar photosphere, by demonstrating that the measured time series differs in amplitude and periodicity when computed on line segments formed at various temperature ranges. From solar simulations on short timescales, we find that convection and its inhibition inside active regions due to strong magnetic fields is responsible for most of the observed effect. This dominant contribution is primarily measured in line segments formed at higher temperature ranges, i.e. deeper parts of the photosphere where convective velocities are expected to be larger and their suppression more noticeable. However, for the lower temperature ranges (high in the photosphere), we detect an effect which does not seem to be explained by convection, and is likely due to a change of the physical conditions. Some indicators allude that the signal in this cooler temperature range could be from the flux effect of active regions (due to its stronger periodogram power at half the solar rotation period). Although, this is not satisfactory supported by the simulated RV flux contributions without considering unexplained lags, we do not rule out that the simulated RV curves might not fully capture line shape variations which affect line segments formed at some temperatures more than other. The explanation by some other, unaccounted physics driving the behavior is outside the scope of this paper, however, we point out to a possible change in the contrast of either the active regions, or the quiet Sun granular pattern \citep[e.g.][]{Janssen&Cauzzi06,Cheung+07}.

With average formation temperature as a diagnostic tool, foreseeable improvements could still be made in its derivation for certain spectral portions. For example, the inclusion of non-LTE departure in the syntheses could enable us to more accurately model the strongest lines, whose line cores are the primary constituents of the coolest temperature ranges. By improving their synthesis and thus also their formation height, one could disentangle their primary source of variability more easily and perhaps better understand the physical processes behind.

We also note that in Sect.~\ref{Sect:4.2}, although we compare the temperature-binned RVs to the \texttt{SOAP-GPU} simulations, which outputs values in agreement with observations when measured across all temperature regimes, it is not straightforward to which extent the comparison is valid when considering only segments of lines. Indeed, even if \texttt{SOAP-GPU} was able to fully model the physics, the simulated velocities would always be averaging over all formation temperatures. Our assumption is that the isolated contributions still affect the entire lines the same way as they would on the line segments where they are the most pronounced, although to a diluted extent when combined with the less sensitive segments. Therefore, despite being unable to compare the magnitudes of the effects, the shapes and phases should be comparable.

It is evident from the existing literature, that it is indeed the interplay between magnetic fields and multi-scale convective motion that is currently impeding RV precision to reach the sub-$\SI{}{\meter/\second}$ level. Although a careful selection of lines result in some temperature bins having higher RV precision, it remains to be determined which partition strategy is the most optimal. The trade-off between the number of bins and the number of lines per bin is so far arbitrarily made and should perhaps be physically motivated by stellar structure models and the instrumental noise level rather than the circumstantial range of temperatures from the sample. In a future paper, we intend to explore this parameter space further by, among other aspects, investigating for which temperature interval a minimum dispersion is attained, and at which break-off temperature the strongest periodogram peak transitions between the fundamental and harmonic periodicities. We will also consider developing a mitigation technique in which a linear combination or gradient of the total RV measured in various temperature ranges can be used to distinguishing the activity signals from a planetary-induced, pure Doppler signal, which should be of equal magnitude and invariant between temperature bins.

\section{Conclusions}\label{Sect:6}

In this paper, we build further upon the technique of measuring radial velocities of individual spectral lines. By employing LTE spectral synthesis, we are able to estimate the average formation temperature of spectral line points as sampled by the HARPS-N and HARPS high-resolution spectrographs, for the Sun and \aCenB{} respectively.

With a careful selection of unblended, symmetric lines and with average formation temperate as a diagnostic tool, we investigate how various activity-related properties depend on the formation temperature. Our primary reasoning being that formation temperature maps more consistently with photospheric depth than line depth, and should be a more suitable variable for probing different layers. We begin by showing that the absolute convective blueshift of line cores becomes a linear function of formation temperature, instead of a non-linear function of line depth. This is observed for both the studied target stars.

We are furthermore able to demonstrate that spectral line segments formed at different temperatures in the photosphere can be used to diagnose the impact of stellar activity on measured RVs, at both rotational (see Figs.~\ref{Fig:6} and \ref{Fig:B2}) and magnetic cycle timescales (see Fig.~\ref{Fig:10}). Comparison with RV simulations covering a few rotations during which the Sun experienced heightened levels of active region coverage, indicates that convection and its inhibition due to magnetic fields is the dominant effect and perturbs spectral parts formed deeper in the photosphere (at higher temperatures). For both short and long timescales, the line segments formed at cooler temperature ranges exhibit an inverted RV variation when compared to the hotter temperature ranges. Measuring the RV for different temperatures in the photosphere could therefore be a powerful tool to differentiate between stellar activity signals and a planetary signal.

It remains to be investigated whether different types of stellar activity can be entirely disentangled through partitioning of formation temperature as a proxy for photospheric depth, and if variations in RV with temperature can be used to mitigate the activity patterns and enhance the signal of potential planetary signals.

\text{}\\
\footnotesize
\noindent
\textit{Acknowledgments.} We thank the referee for useful comments which helped improve the clarity of the manuscript. We thank Lucia Kleint for interesting discussions. This work has been carried out within the framework of the NCCR PlanetS supported by the Swiss National Science Foundation. This project has received funding from the European Research Council (ERC) under the European Union’s Horizon 2020 research and innovation program (grant agreement SCORE No. 851555). This work has made use of the VALD database, operated at Uppsala University, the Institute of Astronomy RAS in Moscow, and the University of Vienna. This work has made use of the HARPS-N solar RVs as well as ESO public HARPS data.

\balance

\bibliographystyle{aa}
\bibliography{References}

\onecolumn
\begin{appendix}

\section{Corner plot for the Sun}\label{Sect:A}

\begin{figure*}[h!]
	\includegraphics[width=0.95\textwidth]{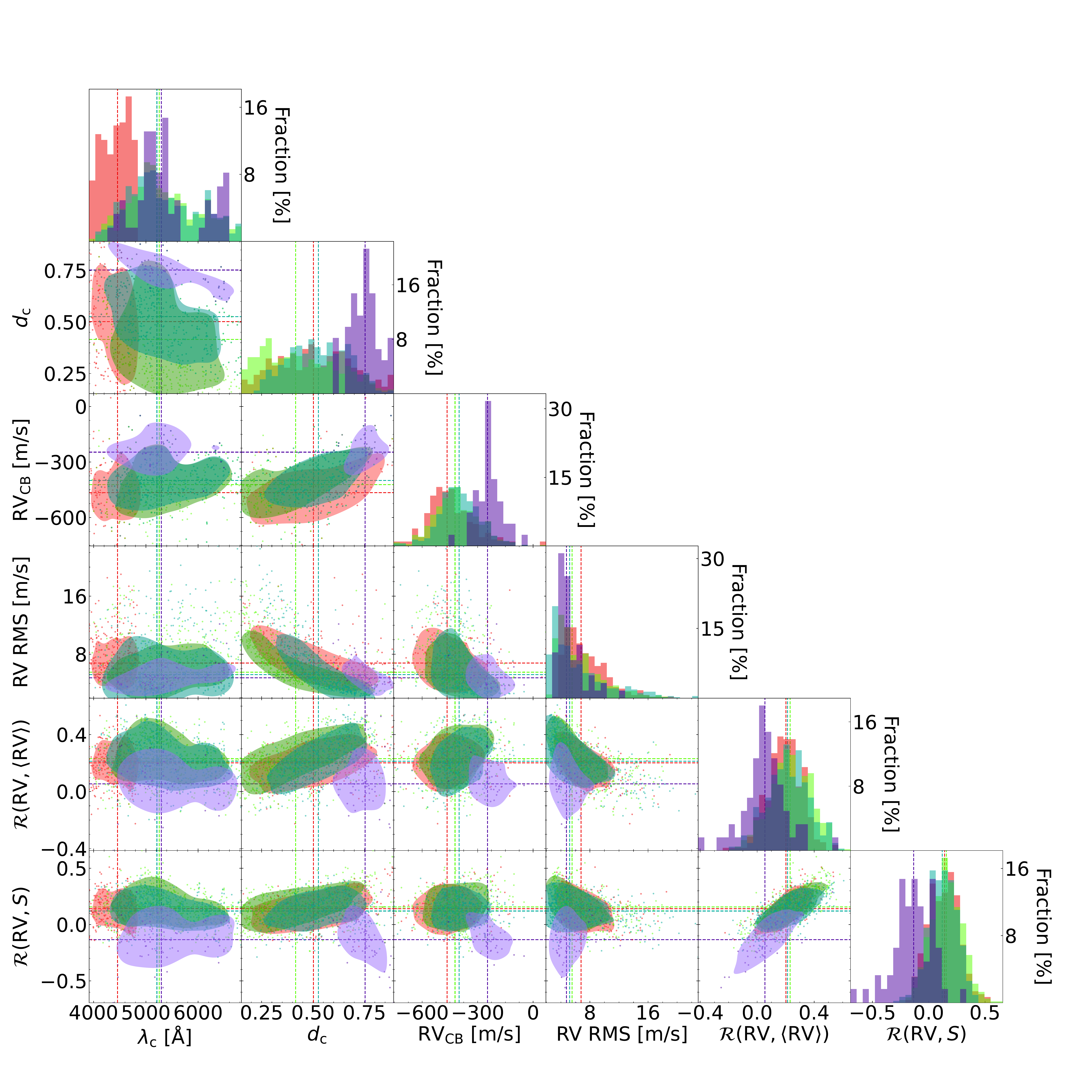}
	\caption{Corner plot of various line parameters of the Sun for $4$ temperature bins. From left to right: central wavelength, line depth, convective blueshift, correlation between LBL RV and mean RV, and correlation between LBL RV and $S$ index. The dashed lines and contours indicate the medians and the ${\pm}1\sigma$ kernel densities, respectively. The color coding is the same as in Fig.~\ref{Fig:6}.}
	\label{Fig:A1}
\end{figure*}

\newpage
\section{Figures and tables for \aCenB{}}\label{Sect:B}

\begin{figure*}[h!]
	\includegraphics[width=\textwidth]{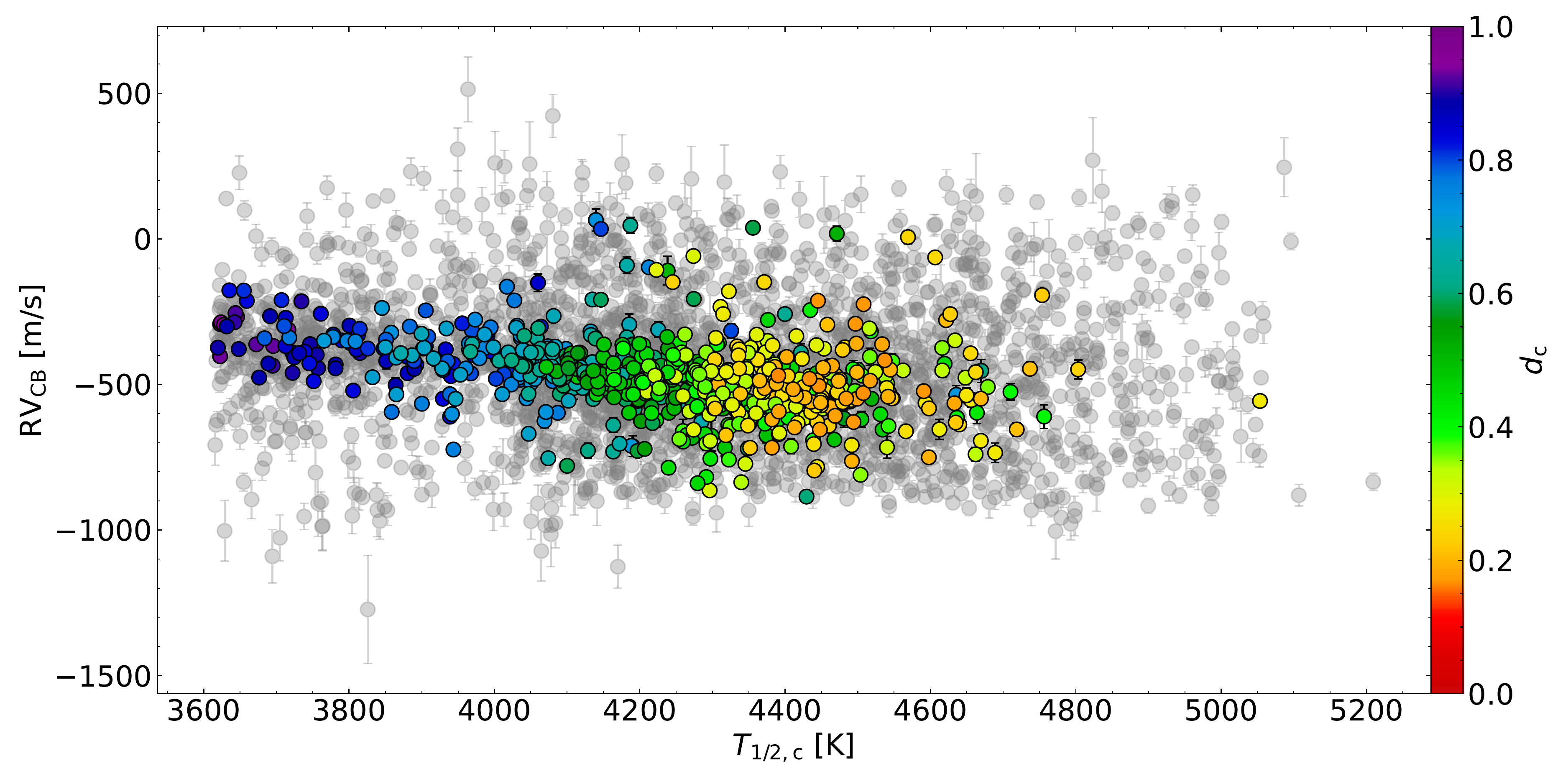}
	\caption{Same as Fig.~\ref{Fig:4}, except for \aCenB{}.}
	\label{Fig:B1}
\end{figure*}

\begin{figure*}[h!]
	\includegraphics[width=\textwidth]{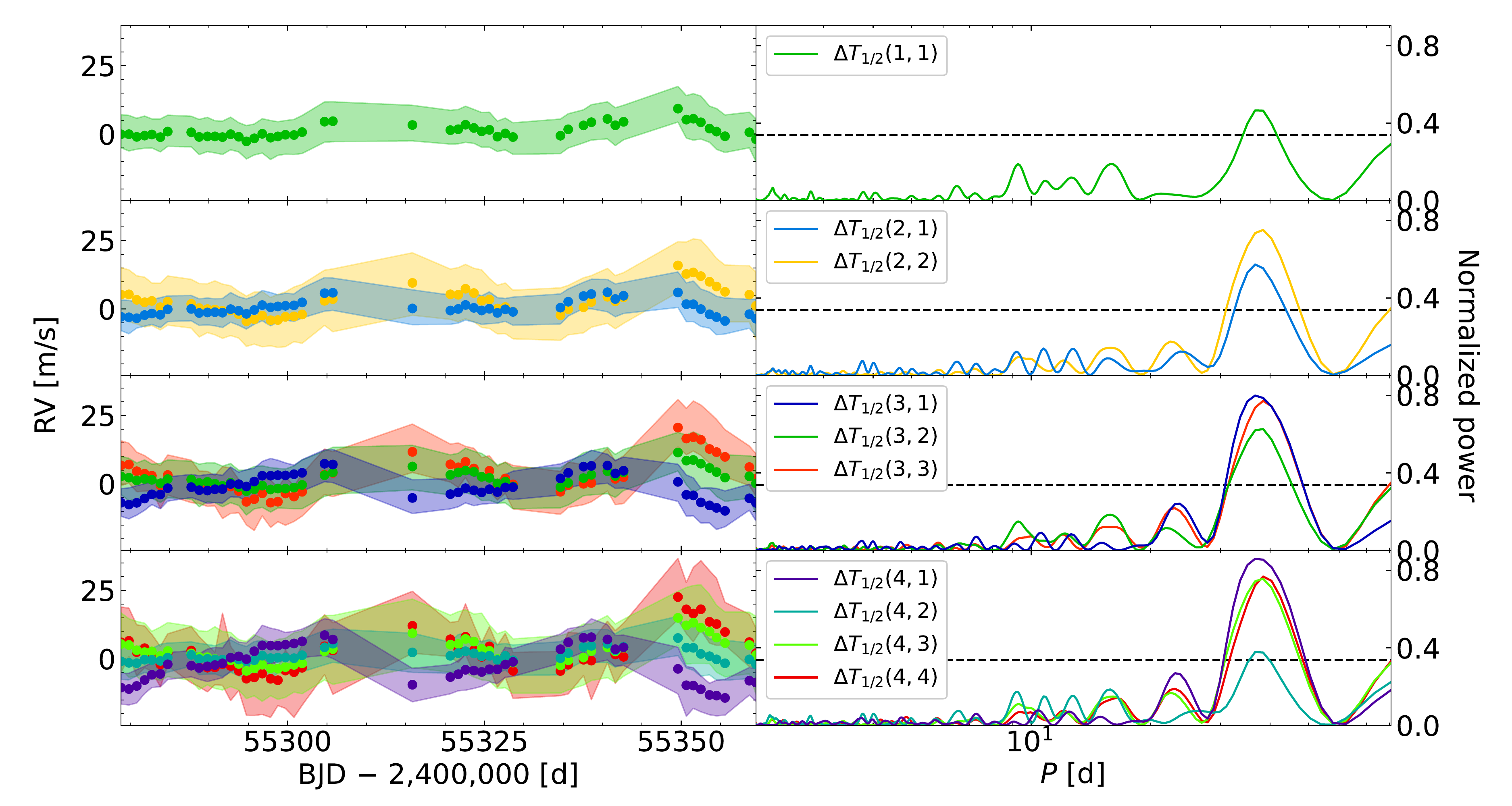}
	\caption{Same as Fig.~\ref{Fig:6}, except for \aCenB{}.}
	\label{Fig:B2}
\end{figure*}

\begin{table*}[t!]
    \caption{Same as Table~\ref{Tab:02}, except for \aCenB{}.}
	\begin{tabular*}{\textwidth}{c @{\extracolsep{\fill}} c @{\extracolsep{\fill}} c @{\extracolsep{\fill}} c @{\extracolsep{\fill}} c @{\extracolsep{\fill}} c}
		\toprule
		\midrule
		\textbf{Nr. of bins} & \textbf{Symbol} & \textbf{$T_{1/2}$ [K]} & \textbf{Nr. of lines} & \textbf{Fraction of lines [\%]} & \textbf{RMS [m/s]}\\
		\midrule
		\multirow{1}{*}{1} & $\Delta T_{1/2}(1,1)$ & 3614-5276 & \multirow{1}{*}{549} & 100 & 2.47\\
		\midrule
		\multirow{2}{*}{2} & $\Delta T_{1/2}(2,1)$ & 3614-4445 & \multirow{2}{*}{468} &  53 & 2.67\\
		                   & $\Delta T_{1/2}(2,2)$ & 4445-5276 &                      &  88 & 4.72\\
		\midrule
		\multirow{3}{*}{3} & $\Delta T_{1/2}(3,1)$ & 3614-4168 & \multirow{3}{*}{470} &  29 & 4.47\\
		                   & $\Delta T_{1/2}(3,2)$ & 4168-4722 &                      &  86 & 3.01\\
		                   & $\Delta T_{1/2}(3,3)$ & 4722-5276 &                      &  15 & 6.36\\
		\midrule
		\multirow{4}{*}{4} & $\Delta T_{1/2}(4,1)$ & 3614-4029 & \multirow{4}{*}{430} &  19 & 6.44\\
		                   & $\Delta T_{1/2}(4,2)$ & 4029-4445 &                      &  48 & 2.11\\
		                   & $\Delta T_{1/2}(4,3)$ & 4445-4860 &                      &  79 & 4.55\\
		                   & $\Delta T_{1/2}(4,4)$ & 4860-5276 &                      &   7 & 7.13\\
		\bottomrule
	\end{tabular*}
	\label{Tab:B1}
\end{table*}

\begin{figure*}[h!]
	\includegraphics[width=0.95\textwidth]{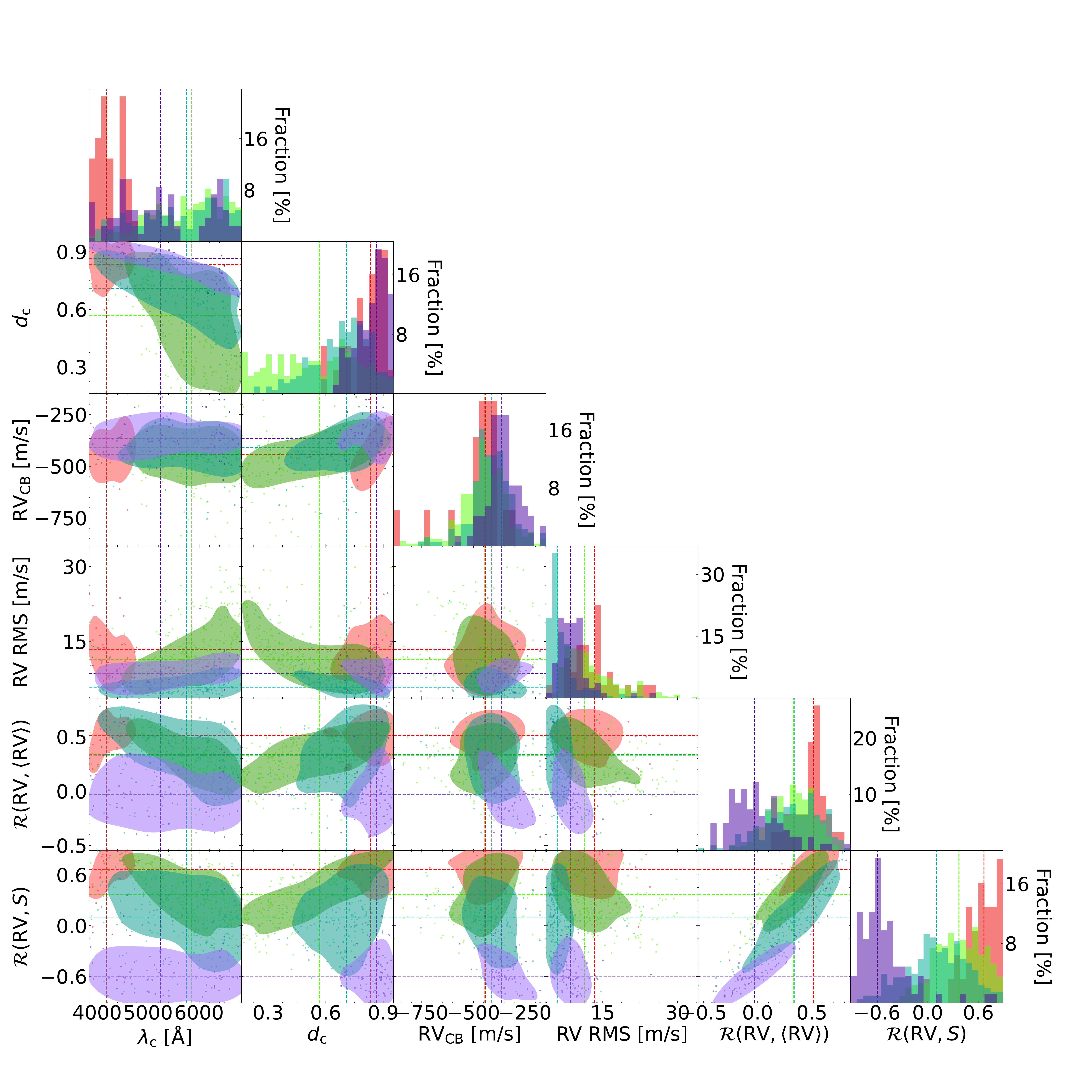}
	\caption{Same as Fig.~\ref{Fig:A1}, except for \aCenB{}.}
	\label{Fig:B3}
\end{figure*}

\end{appendix}

\end{document}